\documentclass[11pt,a4paper]{article}
\usepackage{amsmath,amssymb,amsfonts,a4wide,graphicx,bm,times,psfrag,wrapfig,sidecap, adjustbox}
\usepackage{cite} \usepackage[colorlinks=true,linkcolor=black,
citecolor=black, urlcolor=black]{hyperref}
\numberwithin{equation}{section} \makeatletter
\let\old@startsection=\@startsection \renewcommand{\@startsection}[6]
{\old@startsection{#1}{#2}{#3}{#4}{#5}{#6\mathversion{bold}}}

\def\[{\left[}
\def\]{\right]}
  \def\bee{\be\begin{aligned}}
 \def\eee{\end{aligned} \ee }
  
\def\defeq{\stackrel{\text{def}}=} \newcommand\re[1]{({\ref{#1}})}
\def\be{\begin{eqnarray} } \def\ee{\end{eqnarray}}

 \def\no{\nonumber}
\def\la{\label}  \def\<{\langle}

 \def\>{\rangle}   \def\p{\partial}
\def\a{\alpha} \def\b{\beta} \def\g{\gamma} \def\e{\epsilon} 
\def\eb{ \bm{\e}}
\def\teb{{  \bm{\tilde{\e}}}}

\def\bbvp{ { \bm{\bar\varphi}}}
\def\bbpsi{ \bm{\bar{ \psi }}}
\def\tbpsi{ \tilde{\bm{\psi }}}
\def\bvp{ {\bm{\vp }}}
\def\bb{\mathbf{b}}\def\bbb{\mathbf{\bar b}}
\def\tw{{\tilde w }}
 \def\s{\sigma} \def\t{\tau} 
   
 \def\vp{ \varphi}
\def\th{\theta}

   \def\CO{{
\mathcal{ O} }} \def\CZ{{ \mathcal{ Z} }}  
  
\def\CE{ {\cal E}} 
\def\CH{{\cal H}}

\def\CG{{\cal G}}  
  \def\CD{{ \cal D}}
 
\def\Tr{{\rm Tr}}

\def\d{\delta}

  \def\bxi{\bm{\xi }}  \def\tbxi{\bm{\tilde \xi }}
 
  \def\bpsi{\bm{\psi }}\def\tbpsi{{\bm{\tilde\psi }}}
  \def\etab{\bm{\eta }}\def\tetab{{\bm{\tilde\eta }}}
\def\phib{{{\bm{\phi}}}} \def\tphib{{{ \bm{\tilde\phi}}}}
\def\JJ{\mathbf{a}}\def\FF{{\mathbf{F}}}
 \def\FFT{\tilde{\mathbf{F}}}
\def\JJT{\mathbf{\bar a}}
 \def\hf{{\textstyle{1\over 2}}}
      \def\W{ W}
                  \def\IZ{\mathbb{Z}}
                    \def\IR{\mathbb{R}}
      \def\({\left(}\def\){\right)}
                  \def\CC{\mathcal{C}}  
                   
   \def\FB{{ {\text{FB}}}}
   \def\FFB{ \mathcal{F}  }
    \def\FFBT{ \tilde{\mathcal{F} } }
 \def\tot{\mathrm{tot}}
   \def\wtot{w_{\tot}}
 \def\CF{ \mathcal{F}}
  \def\HH{\mathbf{H}}

   \def\cyl{ \text{cyl}}
\def\tor{\text{tor}}
\def\pp{{_{^{[+]}}}}
\def\mm{{_{^{[-]}}}}

\begin{document}

  \thispagestyle{empty}

\begin{flushright} 
\end{flushright}

\vspace{1cm}
\setcounter{footnote}{0}

\begin{center}

	{ \Large\bf Two-dimensional massive integrable models on a torus}

\vspace{20mm} 

Ivan Kostov
  
 \bigskip

		 {\it  Institut de
		 physique th\'eorique, Universit\'e Paris-Saclay, CNRS and CEA, \\
	    91191, Gif-sur-Yvette, France
	      } 

\end{center}

\vskip18mm \noindent{ The finite-volume thermodynamics of a massive
integrable QFT is described in terms of a grand canonical ensemble of
loops immersed in a torus and interacting through scattering factors
associated with their intersections.  The path integral of the loops
is evaluated explicitly after decoupling the pairwise interactions by
a Hubbard-Stratonovich transformation.  The HS fields are holomorphic
fields depending on the rapidity and can be expanded in elementary
oscillators.  The torus partition function is expressed as certain
expectation value in the Fock space of these oscillators.  In the
limit where one of the periods of the torus becomes asymptotically
large, the effective field theory becomes mean field type.  The mean
field describes the infinite-volume thermodynamics which is solved by
the Thermodynamical Bethe Ansatz.}

 \vskip 4cm
    
\newpage
\setcounter{footnote}{0}

\tableofcontents

\newpage

\section{Introduction}
 
A wide variety of integrable systems in 1+1 dimensions are described
in terms of a factorised scattering \cite{Zamolodchikov:1977jq}.  The
scattering theory provides a very general scheme which applies also
for quantum field theories (QFTs) without Lagrangian formulation.
Remarkably, the scattering theory, although being defined for
asymptotic states in infinite spacetime, can be adjusted to study QFTs
with compactified space or time dimensions.  It is known since long
time \cite{BETH1937915,PhysRev.187.345} that the infinite-volume
thermodynamics of a massive QFT can be expressed in terms of its
S-matrix only.  In 1+1 dimensions, the infinite-volume thermodynamics
can be solved exactly by the Thermodynamical Bethe Ansatz (TBA)
technique introduced by Yang and Yang \cite{YY}.  In the early 90's,
Alexey Zamolodchikov
\cite{Zamolodchikov:1989cf,Zamolodchikov:1991vx,Zamolodchikov:1991vg},
showed how to use the TBA to compute exactly the finite-volume
ground-state energy at zero temperature.  Since then a lot of progress
has been made in computing the finite volume/temperature effects in
integrable models, including some correlation functions.  For a review
see e.g \cite{Mussardo-Book}.
  
There are reasons to expect that it is possible to formulate also the
{\it finite-volume} thermodynamics of a massive solvable QFT solely in
terms of its scattering data.  Indeed, the energies of the excited
states in finite volume can be determined, at least in principle, by
analytical continuation of the TBA solution \cite{Dorey:1996re} which
in turn is determined by the S-matrix.  Once the full spectrum of the
finite-volume Hamiltonian is known, the partition function of the
Euclidean QFT on the torus is known as well.

On the other hand, the diagonalisation of the finite-volume
Hamiltonian seems, at least at present, to be an extremely difficult
task.\footnote{In the massless limit, this is possible in many cases
due to the conformal symmetry which, together with the modular
invariance, was sufficient to determine the torus partition functions
\cite{ITZYKSON1986580,diFrancesco1987,Kostov:1988pt,Foda:1988in,Saleur:1987pk}.}
The only known partition functions of massive theories on the torus
are those of the so called generalised free theories characterised by
constant S-matrices \cite{Saleur1987,Klassen:1990dx}.\footnote{The
partition functions of some lattice models on tori with special
geometry were recently computed in
\cite{Jacobsen:2018pjt,Bohm:2022ata,Bajnok:2020xoz}.  Unfortunately
these results cannot be used in the continuum limit.  }
 
In this paper, I propose an alternative formulation of the finite-size
thermodynamics, for which no knowledge of the finite-volume energy
spectrum is necessary.  The formulation in question concerns the
simplest case of a theory of one single neutral particle and without
bound states in the spectrum.  The principal claim is that in a theory
with factorised scattering, the torus partition function can be mapped
onto the grand canonical ensemble of relativistic loops embedded in
the torus and interacting through scattering factors associated with
the crossings.  In the loop gas description, the finite-volume effects
are due to non-contractible loops winding around the time and space
cycles.
  
The advantage of the loop-gas formulation is that one can accomplish a
separation, in the spirit of \cite{PhysRev.187.345}, of the dynamical
part, which is expressed in terms of the S-matrix, from the
statistical part, which appears through the path integral for the
loops.  This is achieved by performing a Hubbard-Stratonovich type
transformation in order to decouple the two-body interaction of the
loops.  The HS auxiliary fields are associated with the two homology
cycles of the torus.  The decoupling makes possible to perform
explicitly the integration over the loops, including the sum over the
winding numbers.  The sum over the loops results in an effective
interaction potential for the HS fields.  In order to generate the
Gaudin measure, an extra pair of Faddeev-Popov ghost fields is added.
  
As a result, the torus partition function is expressed as certain
expectation value in an effective quantum field theory (EFT) for a
pair of bosonic and fermionic oscillators defined in the complex
rapidity plane, with two-point function determined by the S-matrix.
In the limit when one of the periods of the torus becomes
asymptotically large, this EFT was formulated in \cite{Kostov:2020aa}.

The text is organised as follows.  In section \ref{section:cylinder},
the partition function of the QFT on a cylinder is formulated as a
loop gas.  The aim of this section is to verify that the free energy
of the loop gas matches the finite-volume vacuum energy computed by
the TBA. First I recall the path integral over loops winding given number of times around the cylinder.  Importantly,
the path integral in question can be expressed in terms of the wave
functions of {\it on-shell} particles propagating in the infinite
Minkowskian spacetime. The free energy of the loop gas, obtained by
summing over all winding numbers, is shown to reproduce correctly the
effective central charges of the massive boson and the massive Majorana fermion.  Then the same derivation is carried out for a
theory with non-trivial scattering with the help of HS transformation.
 The outcome is the EFT formulated in
\cite{Kostov:2020aa}, which has been shown to be equivalent to the
TBA.

In section \ref{sec:EQFTtorus}, the torus partition functions of the
generalised free theories, the free massive boson (FB) and the Ising
Field Theory (IFT), are formulated as ensembles of loops immersed in
the torus.  While the partition function of the free boson is the
exponential of the path integral for one loop, this is not the case
for the IFT, because the loops interact through minus signs associated
with their crossings.  Although this is aside of the main goal of this
paper, in order to achieve maximal clarity I give a combinatorial
derivation of the expression of the modular invariant partition
function as a sum of four building blocks found by Itzykson and Saleur
\cite{Saleur1987} and Klassen and Melzer \cite{Klassen:1990dx}, based
on the loop gas.
Finally, in section \ref{sec:WindIntTheor} I  formulate the main result
of this paper, the effective field theory for
a massive integrable QFT on a torus, 
with the sinh-Gordon model as an example.  
The Feynman graph expansion for the EFT 
is formulated  in appendix  \ref{sec:Feynman}.

   \section{Integrable QFT on an infinite  cylinder}
   \la{section:cylinder}

In this section, the mapping of the integrable QFT to a loop gas will
be studied in detail for the simpler and well studied limit  of  the $L\times R$  torus in which   the $R$-cycle is taken asymptotically large,
{\it i.e.}  when all exponential corrections in $R$ are neglected.
This limit  describes a QFT in infinite space at finite temperature $\sim 1/L$, with  IR cutoff introduced by imposing  periodic boundary conditions at distance $R$. 
 In the Hamiltonian formulation with the time direction along the $L$-circle, 
the  partition function  reads
\be
\la{ThermalPF}
\CZ^{(L)}=  \Tr_{_{\! R}}  \  e^{- L \CH}
\ee
where $\CH$ is the infinite-volume Hamiltonian and $\Tr_{_{R}}$ 
is the trace in the Hilbert space associated with the  large 
space circle $R$.   
  On the other hand, if the space is associated with  the $L$-circle,
   the partition function \re{ThermalPF} determines the 
  finite-volume vacuum energy 
\be
\la{vacenergy}
 \CE(L) =-  \lim_{R\to\infty} { \log \CZ^{(L)}  \over  R}.
\ee

The aim of this section is to express the finite-volume vacuum  energy $\CE(L)$ as the free energy in a  statistical ensemble 
loops embedded in the cylinder. In a free theory, the vacuum energy 
\re{vacenergy} is given by the path integral of a single loop.
In a  theory with non-trivial factorised scattering,  the loops  experience
pairwise interaction. In this case the  free energy of the loop gas   is obtained by summing up the cumulant expansion, which is  another way to derive the TBA  equations.

%
%
%
%
%
%

    \subsection{Path integral for a loop on a cylinder}           
  \label{sec:periodicwrapping}
 
 In order to develop the loop-gas formalism, let us  work out  the 
 path integral for a single loop.  
The path integral $\CF^{(L )}$ for a loop immersed in the
cylinder is a sum over topological sectors characterised by the number
$w$ of times the loop coils around the cylinder,
 \be \la{sumovertops} \CF^{(L )} = \sum_{w\in\IZ} [\CF^{(L )}]_{w }.
 \ee
 The winding number can take both positive and negative values
 depending on the orientation and two winding numbers add
 algebraically.
 
\begin{figure}[h!]
            \centering
			\includegraphics[width=14.5 cm]{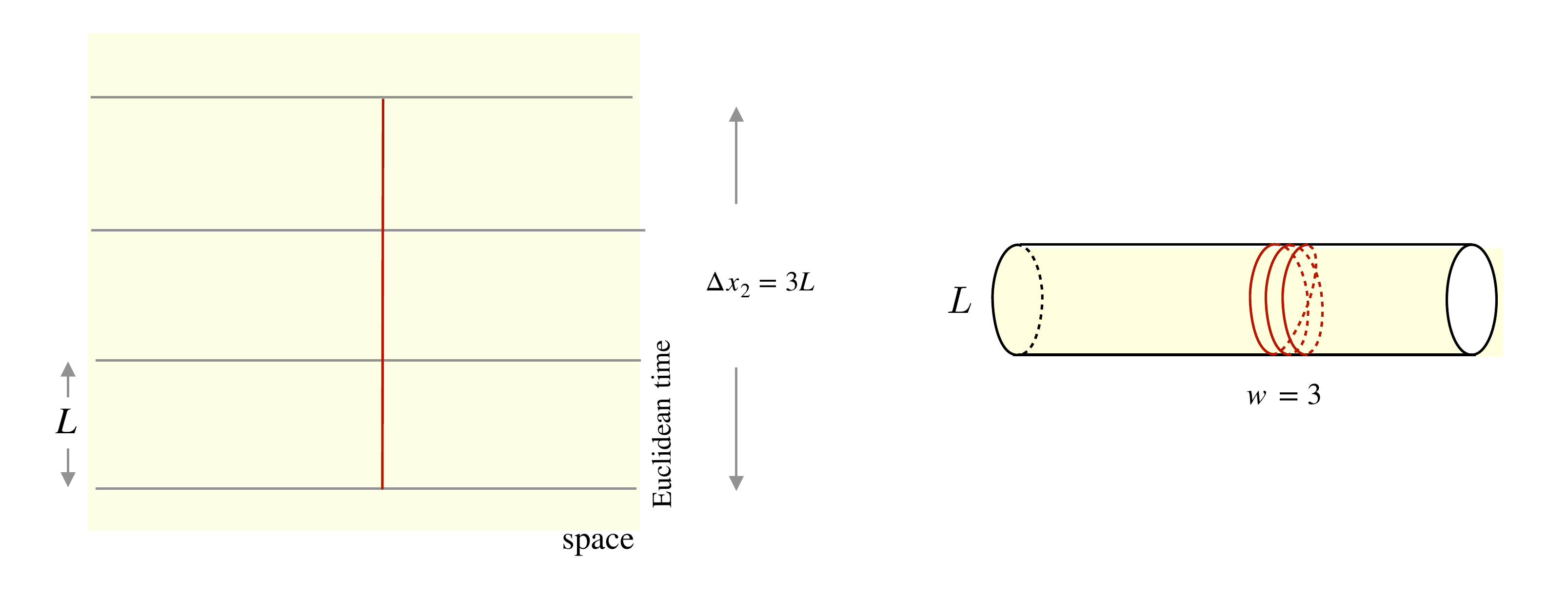}
\vskip -0.5cm \caption{ \small A loop with winding number $w $ as a
path integral with a discontinuity $\delta x =\{ 0, w L\}$ for $w=3$ }
  \label{fig:fig2x}
 \end{figure}

 Let me   sketch of the  computation of  $ [\CF ^{ (L)}]_w $. For that 
 I will use the obvious  relation between the 
 integration measure for loops   on the cylinder 
 having given winding number and  that the loops in the {\it infinite} 
 Euclidean spacetime with  inserted  discontinuity, as illustrated in
fig.  \ref{fig:fig2x},

Consider  
the path integral (per unit volume) for a loop  fluctuating in the {infinite} Euclidean space,  
 with inserted  discontinuity $\delta
\vec x = \{ \delta x_1, \delta x_2\}$.
Denote this path integral by  $\CF( \delta \vec x)$.
The path integral in the sector with winding number $w$ 
is obtained by choosing $\d x_1=0$ and $\d x_2 = w L$,
\be \la{Fwww} [\CF ^{ (L)}]_w =RL\  \CF( \delta \vec x)\Big|_{\delta \vec
x = \{ 0, w L\}}\, . 
 \ee

   A derivation of the path integral measure
for open relativistic loops in infinite Euclidean space is given, for example, in Polyakov's book
\cite{Polyakov:1987ez}.  With small adjustments, the derivation there can be  applied  for closed loops as well.
   It is however  simpler to  manipulate   the expression for  
 the propagator of a massive scalar relativistic
particle in  two-dimensional Euclidean (infinite) spacetime
  \bee \la{somewn0} \CG( \delta \vec x ) &= \int {d^2 k\over
  (2\pi)^2} \ {e^{ i \vec k \cdot \delta \vec x} \over \vec k^2 +m^2 }
  = \int {d^2 k\over (2\pi)^2} \int _{ }^{\infty}{d\tau } e^{- ( \vec
  k ^2 + m^2) \t + i \vec k\cdot \delta \vec x}  
  .  \eee
The integrand in the second representation
can be interpreted as  the Boltzmann amplitude for 
   loops with 
  a marked point,
 proper time $\t$ along the 
loop, 2-momentum $\vec k$,
  and    discontinuity $\vec x(\t)-\vec x(0)= \d\vec x$ 
  at the marked point. 
The integral \re{somewn0}  is not yet $\CF(\d\vec x)$. To obtain
$\CF(\delta \vec x)$, one should undo the marked point
by  modifying 
the integration measure over the proper time as $ d\t\ \to\ d\t\to d\t/\t$,
 and add an  extra  factor of $1/2$  since the loop is assumed  non-oriented.
The resulting integral reads
  %
  \bee \la{somewn1} \mathcal{F} (\delta  \vec x ) &= \hf \,  \int
  {d^2 k\over (2\pi)^2} \int _{ }^{\infty}{d\tau \over \tau} e^{- (
  \vec k ^2 + m^2) \t + i \vec k\cdot \delta  \vec x} \\
&=- \hf  \int {d^2 k\over (2\pi)^2} \log\( \vec k^2 +m^2 \) e^{ i
\vec k \cdot \delta \vec x }
 .
\eee
%
 
%

If $\delta x_2\ne 0$, the integral with respect to $k_2$ can be taken
by residues and the integrand takes the form of the wave function of
on-mass-shell particle with momentum $k_1$, analytically continued to
imaginary time $t= - i \delta x_2$,
\bee \la{rlpathintegral-bis-cyl} \mathcal{F} (\delta \vec x) & =
{\textstyle{1\over 2}} \, RL\, {1 \over |\delta x_2|} \
\int_\mathbb{R} {d k_1\over 2\pi} \ e^{ - \sqrt{k_1^2+m^2}| \delta
x_2|+ i k_1 \delta x_1 } \\
   & = {\textstyle{1\over 2}} \, RL\, {1 \over |\delta x_2|} \
   \int_\mathbb{R} {d p(\th)\over 2\pi} \ e^{ - E(\th)| \delta x_2|+ i
   p(\th) \delta x_1 } \qquad (\delta x_2\ne 0) .  \eee
In the last line,  the integral is written in terms of the rapidity
$\th$ which parametrises the mass shell $E^2-p^2 = m^2$,
\be \la{rlpathintegral-bis} E(\th)= m\cosh\th, \ \ p(\th) =m\sinh\th.
\ee
 Thus for the path integral in the sector with non-zero winding number
$w$ one can write the integral representation
	 \bee \la{windloopintegr} [\CF^{(L)}]_w = \mathcal{F} _ {w },
	 \quad \mathcal{F} _ {w } \defeq { \hf } \int _{\mathbb{R} }
	 {Rdp(\th )\over 2\pi} \ { e^{ -| w| L E(\th ) } \over |w|}\
	 \qquad ( w\ne 0).  \eee

\bigskip

\bigskip

\noindent

As for the path integral $[\CF^{(L)}]_0$ for a contractible loop ($w=0$),
it is of the form 
\be 
[\CF^{(L)}]_0 =-RL \e_0, 
\ee
where  $\e_0$   is the infinite-volume energy
density of the vacuum.   The computation of this quantity
requires an UV cutoff because the path integral is dominated by small loops.
  Since the focus here is on the finite-volume effects, a normalisation $\e_0=0$ will be assumed for the moment.

   \subsection{ Free massive boson }
   \la{section:FMBonacbos}
   
The free massive boson is defined by the Euclidean path integral
   \bee \la{partffreeb} \mathcal{Z} _{ \mathrm{FB}} ^{ (L)} = \int \CD\vp \
   \exp\left[ \int d^2x\ \vp( \nabla^2-m^2)\vp\right] \eee
 with periodic boundary condition $ \vp(x_1,x_2+L ) = \vp(x_1, x_2) $.
 The free energy\footnote{ Here it is convenient to include the
 inverse temperature in the definition of the free energy.  }
 $\mathcal{F}_{\mathrm{FB}} ^{ (L)} \equiv \log \mathcal{Z}
 _{\mathrm{FB}} ^{ (L)} $ is that of a gas of free oscillators at
 temperature $1/L$.  With the normalisation $\e_0=0$,
\bee \la{somewn} \mathcal{F}_{\mathrm{FB}} ^{ (L)} & = - R \int
_\mathbb{R} { d p \over 2\pi} \ \log\left( 1- e^{- L E_p}\right) ,
\qquad E_p\equiv \sqrt{p^2+m^2} \\
   &= R \ {\pi \over 6 L} \ c _0 (mL).  \eee
The scaling function $ c _0( mL) $, known as {\it effective central
charge}, tends in the limit $L m\to 0$ to the central charge of the
Virasoro algebra for a free neutral massless boson,
\be c _0 (mL) \underset{ {m=0} }\to - {6 L\over \pi} \int _\mathbb{R}
{dp\over 2\pi} \log(1- e^{- L |p|}) = 1\, .  \ee
In the opposite limit, $mL\to\infty$, the effective central charge
vanishes.

Now let us obtain the rhs of \re{somewn} by evaluating the free energy
of the grand canonical ensemble of loops embedded in the cylinder.
Since the loops do not interact, the free energy\footnote{The winding
numbers can have both signs because the loop can wrap the cylinder in
both directions.  The path integral depends only on the absolute value
of $w$, hence the factor $2$ on the rhs.} is equal to the path
integral for a single loop.  The path integral contains a sum over all
winding numbers
 \bee \la{somewn01} \mathcal{F} ^{ (L)} & =\sum_{ w \in\IZ} \mathcal{F}
 _{ w} ^{ (L)}\, =
 2 \sum_{w=1}^\infty \mathcal{F} _{w} ^{ (L)}.  \eee
After inserting the representation \re{windloopintegr}, the sum over
the winding numbers can be done explicitly, with the result
   \bee \la{somewnbis} \mathcal{F}_{\mathrm{FB}} ^{ (L)} &= - \int
   _\mathbb{R} { R d p(\th)\over 2\pi} \ \log\left( 1- e^{- L
   E(\th)}\right) .  \eee
 This is exactly the free energy of the free massive boson, eq.
 \re{somewn}.

   \subsection{ Free massive Majorana fermion }
   \la{section:FMBonacylinfer}

In the case of a free massive Majorana fermion, the only new element
is the sign factor associated with the intersections of loops.  Since
any two loops winding around the cylinder intersect each other an even
number of times, their intersections do not produce signs.  However a
loop with winding number $w$ intersects itself $w-1$ times, as
illustrated for $w =3$ by fig.  \ref{fig:fig2x}, hence an extra factor
$(-1)^{w-1}$.  Taking the signs into account, sum over the winding
numbers gives
\bee \la{somewn1m} \mathcal{F}_{\mathrm{FF}} ^{(L)} &= 2
\sum_{w=1}^\infty (-1)^{w-1} \mathcal{F} _w \\
  &= R \int _\mathbb{R} {d p(\th)\over 2\pi} \ \log\left( 1+ e^{- L
  E(\th)}\right) \equiv {\pi \over 6} {R\over L} \ c _{1/2} (mL) .
  \eee
The effective central charge $c _{ 1/2}(mL) $ interpolates between
$1/2$ in the UV limit and $0$ in the deep IR limit.

The free boson and fermion can be considered as scattering theories
with constant scattering factor $\s=1$ for the boson and $\s= -1$ for
the fermion.  The expression for the free energy for such a
`generalised free theory' is
\bee \la{somewn12} \mathcal{F}^{(L)}_{\mathrm{FB,FF}} &= 2
\sum_{w=1}^\infty \s^{w-1} \mathcal{F} _w^{(L)} \\
   & = -\s\ \int _\mathbb{R} {d\phi(\th) \over 2\pi} \ \log\left(
   1-\s\, e^{- L E(\th)}\right) ,\qquad \phi(\th) \equiv R \, p(\th).
   \eee
The integration measure  turns out to be the flat measure for the phase $\phi$
acquired by the one-particle wave function after a tour around the
space circle.\footnote{The standard argument is that  the free energy is
a discrete sum  of the rapidities satisfying  the quantisation condition
$\phi(\th) \in 2\pi \IZ$. 
Since $R$ is assumed asymptotically large, the exponential corrections
due to the discreteness of the spectrum can be neglected and the sum
over equi-distant phases can be replaced by an integral with measure $d\phi/2\pi$.
However the loop gas formulation does not  involve discrete momenta. 
The measure in \re{somewn12} follows from the path integral for the  winding particles
and does not use any discretisation.
}

   \subsection{ Interacting  massive integrable QFT }
\la{section:loopgascyl}

Now we are prepared to formulate the loop gas for an interacting
integrable QFT. In a relativistic integrable QFT with one particle
species and no bound states, the scattering matrix
$S(p_1,p_2)=S(\th_1-\th_2)$ is a phase factor satisfying the
requirements of real analyticity, unitarity and crossing
\cite{Zamolodchikov:1977jq},
\bee \la{proprsS} &S(\th)^* = S(- \th ^*) \qquad\qquad \qquad
\text{(real analyticity)} \\
  &S(\th)S(-\th )=1 \qquad\qquad \qquad \qquad \text{(unitarity)} \\
  &S(\th )= S(i\pi - \th) \qquad\qquad \quad \text{(crossing
  symmetry)} .  \eee
The TBA statistics, bosonic or fermionic, of the particles with the
same momenta is determined by the sign of the scattering factor at
$\th=0$,
\bee \s \equiv S(0) =\pm 1.  \eee
In the original Yang-Yang paper \cite{YY} it is assumed that $\s=-1$,
but in theories with one space dimension the statistics of identical
particles is physically irrelevant because it mixes up with the
interaction.  It was shown by Wadati \cite{doi:10.1143/JPSJ.54.3727}
that changing both the statistics and the interaction, one can obtain
a different description of the same integrable model.  Since bosonic
S-matrices are known to describe interesting physical phenomena
\cite{Mussardo:1999aj,Cordova:2021fnr}, both TBA statistics will be
discussed here.

Once the path integral over loops is expressed in terms of on-shell
wave functions, eq.  \re{rlpathintegral-bis-cyl}, the loop gas
description developed in the previous subsections can be extended to a
theory with nontrivial scattering by the following recipe:
 { Intersections of tho segments of
loops with rapidities $\theta_1$ and $\theta_2$ is counted with a
factor $S(\th_1-\th_2)$.}
  
   \smallskip

Scattering factors are associated as well with the self-intersections
of a loop.  In this case the two segments have coinciding rapidities
and the extra factor for a self-intersection is $\s =S(0)$.

As any two loops winding around the cylinder intersect an even number
of times, the scattering factors compensate thanks to the unitarity
property \re{proprsS} of the S-matrix.  Therefore the Boltzmann weight
for a configuration of $N$ loops with rapidities $\th_1,..., \th_N$
factorises into a product of $N$ one-loop Boltzmann weights given by
the integrand in \re{windloopintegr}.  However the partition function
does not exponentiate trivially as in the (generalised) free theory,
eq.  \re{somewn12}, because the rapidities get entangled through the
integration measure.

   \begin{figure}[h!]
            \centering
            \includegraphics[width=8.5 cm]{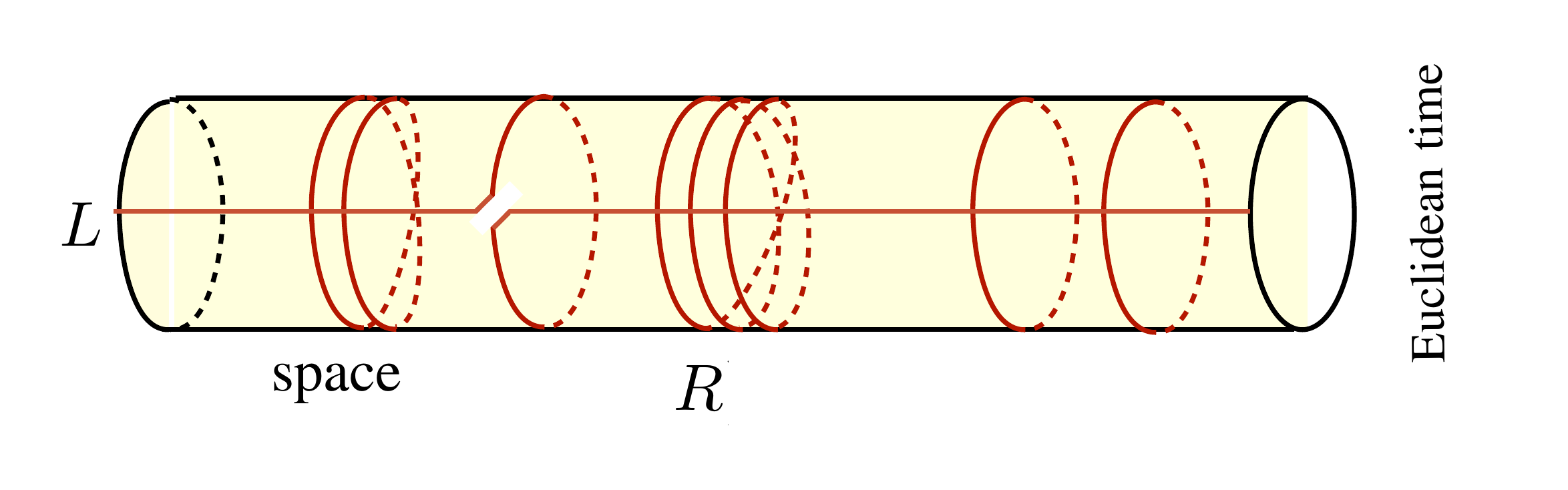}
\caption{ \small A winding particle touring around the world.  }
  \label{fig:figcilinder1}
 \end{figure}

  As already mentioned, the measure in the integral over the rapidity
  is the flat measure for the phase shift $\phi = R d p(\th)$ of the
  semiclassical wave function for distance $R$.  The derivative of the
  phase is proportional to the density of the states in the
  Hamiltonian description.  In an interacting theory, the periodicity
  condition in $R$  becomes more complicated since   the phase shift
  of the wave function of a particle takes contributions from the
  scattering with all other particles in the ensemble.  One can
  associate this phase shift with a loop making an extra `tour around
  the world', as shown in fig \ref{fig:figcilinder1}.  The phase
  acquired after such a trip gets dressed by the scattering factors
  from crossing the other loops,
  \be \la{dressingC} \phi(\theta| \th_1,..., \th_N ) =Rp(\theta )+i
  \sum_{k=1}^N |w_{k}| \log S(\theta_k -\theta).  \ee
 Because of the dressing, the integration measure for $N$ loops does
 not factorise.  The flat measure with respect to the phases
 $\phi_1,..., \phi_N$ of the $N$ loops,
  \bee \la{Nphases} \phi_ j & =Rp(\th_j)+i \sum_{k=1}^N |w_{k} | \log
  S(\th_k-\th_{j}) , \ \ \ j=1,..., N, \eee
contains the Jacobian (known as Gaudin determinant
\cite{PhysRevD.23.417}) for the change of variables from phases to
rapidities,
\bee \la{GaudinD11} \prod_{j =1}^N {d \phi_j \over 2\pi } = \prod_{j
=1}^N {d\th_j \over 2\pi } \ \det \left( {\p\phi_j\over\p \th_{k} }
\right) .  \eee

   \subsection{ Hubbard-Stratonovich fields}
\la{gaussrepcyl}

The presence of a Jacobian in the integration measure \re{GaudinD11}
has the effect that the free energy becomes a sum of clusters of loops
with the structure of branched trees
\cite{Kostov:2018ckg,Kostov:2018dmi}.  The generating function for the
sum over trees solves the TBA equation.  A field-theoretical
derivation of the exact cluster expansion was given in
\cite{Kostov:2020aa}.
 
In the context of the loop gas on a cylinder, the effective field
theory of \cite{Kostov:2020aa} arises after a HS type transformation
\cite{SHtrsfmtn,HStransftn} which decouples the two-body interaction
of loops.  As a result, the ensemble of interacting loops is
reformulated as an ensemble of independent loops interacting with a
pair of auxiliary fluctuating fields.  In order to decouple the
interactions, replace in the loop amplitudes
   \be \la{replacnemt} \phi(\th)\equiv Rp(\th)\ \to\ \phib(\th),
   \qquad L E(\th)\ \to\ \eb(\th), \ee
    where    $\eb (\th)$ and $\phib (\th)$ 
    are  gaussian holomorphic  fields  with connected   correlation function
 \bee \la{correlscy} \langle \phib ( \theta ) \eb (\theta ') \rangle
 _c = i \log S( \theta -\theta '), \quad \< \eb(\th)\eb(\th')\> _c=
 \<\phib(\th)\phib(\th')\>_c=0, \eee
designed to generate the dressing of the phase in \re{dressingC}.
According to \re{replacnemt}, the two gaussian fields should be given
classical values
   \bee \la{defgf} \langle \eb (\theta)\rangle = L E(\theta),\ \ \
   \langle \phib (\theta)\rangle = R p(\theta) . 
   \eee
The path integral over the loops with winding number $w$, eq.
\re{windloopintegr}, is now replaced by an operator,
 \bee \la{OpFcyl} \CF_w^{(L)} &= { 1\over 2 } \, \int _\mathbb{R} {R d
 p(\th)\over 2\pi} \ {e^{- |w| L E(\th)}\over |w|}\ \ \to \ \ \FF_w =
 { 1\over 2 } \, \int _\mathbb{R} {d\phib(\th)\over 2\pi}\ {e^{- |w|
 \eb(\th)}\over |w|} \qquad (w\ne 0).  \eee
The `operator differential' $d\phib(\th )$ has no precise mathematical
meaning, but can be given an operational definition as the operator
whose expectation value is the differential of the expectation value
of the operator $\phib(\th)$.  With this definition, the expectation
value
	\bee \la{ZNNonecy} \left\langle \prod_{j =1}^N e^{-|w_j|\eb
	(\th_j) } d\phib ( \th_j ) \right\rangle = \prod_{j =1}^N
	e^{-|w_j| L E (\th_j) } \prod_{j =1}^N {d\phi_j \over 2\pi } \eee
generates the Gaudin measure given by the product of the differentials
on the lhs of eq.  \re{GaudinD11}.

Now the sum over the winding numbers can be performed explicitly,
resulting in a remarkably simple operator representation for the
partition function,
   \bee \la{operccyl} \CZ_\cyl^{(L)} = \< e^{\FF _{\cyl }}\>, \qquad
   \mathbf{F_\cyl} = - \sigma \int _\mathbb{R} {d \phib (\theta)\over
   2\pi} \ \log\left( 1-\sigma e^{- \eb (\theta)}\right), \eee
with the dependence on $L$ coming from the bare expectation value of
$\eb(\th)$, eq.  \re{defgf}.

However, the simplicity of this expression is deceiving because of the
operator differential $d\phib(\th)$ which requires a precise
prescription for calculating the expectation value.  Namely, to
evaluate the partition function one should first expand the
exponential of the free energy operator $\FF$, and then apply
\re{ZNNonecy} term by term.  In other words, the evaluation of the
partition function \re{operccyl} brings us back to the original sum
over interacting loops.

A more manuable operator representation can be constructed by bringing
into the game an extra pair of fermionic fields $\bpsi(\th)$ and
$\etab(\th)$ with the same correlation as the bosonic pair,
   \be \la{defetapsi} \langle \bxi (\th ) \etab (\th')\rangle = i \log
   S( \th-\th') , \qquad \< \etab(\th) \etab(\th')\> =\<
   \bxi(\th)\bxi(\th')\> =0.  \ee
With the help of the fermionic fields one can amend the operator
representation \re{ZNNonecy} so that on the rhs the Gaudin determinant
is generated automatically, namely
 \be \la{operepsum} \begin{aligned} \left\langle \prod_{j } e^{- |w_j|
 \eb (\th _j)} \left[ \p \phib (\th_j) -|w _j| \, { \etab}(\th_j) \p
 {\bxi} ( \th_j) \right] \right\rangle &= \prod_{j =1}^N e^{-|w_j| L E
 (\th_j) } \ \det \left( {\p\phi_j\over\p \th_{k} } \right) .
  \end{aligned}
  \ee
The proof of this remarkable identity is given in appendix B of
\cite{Kostov:2018dmi}.
Again, the sum over the winding numbers can be performed explicitly,
resulting, together with \re{correlscy} and \re{defgf}, in the
operator representation proposed in \cite{Kostov:2020aa},\footnote{
The connection with the notations used in \cite{Kostov:2020aa} is
$\{\eb, \phib , \etab, \bpsi\} \leftrightarrow \{ \bar\bvp,
\bvp,\bar\bpsi, \bpsi\} $.  }
\bee \la{partfGENbisTBA} \CZ_\cyl^{(L)}&= \< e^{\FF_\cyl}\>, \qquad
\mathbf{F}_\cyl = -\s \int_{\mathbb{R} }{d\theta \over 2\pi }\ \left[
\log\left( 1 -\s e^{- \eb }\right) \partial _\th \phib + { \etab
\partial _\th \bxi \over 1-\s\, e^{ {\eb } } } \right] .  \eee
By obvious reasons, I will refer to $\FF_\cyl$ as the operator of the
free energy.  The expectation value \re{partfGENbisTBA} generates, as
shown in ref.  \cite{Kostov:2020aa}, the exact cluster expansion
obtained in \cite{Kostov:2018ckg,Kostov:2018dmi}.

\subsection{Ward identities  }

Let us see how the TBA equation appears.  First notice that, since the
interaction potential $\mathbf{F}_\cyl $ is linear in the field
$\phib$, the field $\eb $ has no dispersion.  With the dressed
expectation value defined as
 \be \la{expvaluedr} \< \CO\> _\cyl \equiv { \left\< \CO \
 e^{\FF_\cyl} \right> \over \left\< e^{\FF_{\cyl}}\right>}, \ee
this means that
\be \la{dispersionlessf} \left\< \eb(\th)\eb(\th') \right>_{\cyl}
=\left\< \eb(\th) \right> _{\cyl}\left\< \eb(\th') \right> _{\cyl}.
\ee
 Furthermore, the fields $\eb$ and $\phib$ satisfy the obvious Ward
 identities
\begin{align}
\la{Widentitya} \left\< \eb(\th) \right> _{_{\cyl}}& = LE(\theta )
+\s\, \int_{-\infty}^\infty {d\theta '\over 2\pi} K(\theta -\theta ')
\left\< \log \(1- \s\, e^{ - \eb (\theta ')}\) \right>_{{\cyl}} \, ,
\\
\left\< \p\phib(\th) \right>_{{\cyl}} & = R\p p(\theta ) +
\int_{-\infty}^\infty {d\theta '\over 2\pi} K(\theta -\theta ')
\left\< { \p \phib (\th')\over e^{ \eb (\theta ')}- \s } + e^{\e(\th')
} { \etab (\theta )\partial \bxi(\theta ) \over (e^{ \eb (\theta ')}-
\s )^2 } \right> _{\cyl} \, , \la{Widentitya2}
\end{align}
where
\bee \la{defscK} K(\theta - \theta ') & =- \left\langle \p \phib (\th)
\eb (\th') \right\rangle_c = - i\partial_\theta \log S(\theta -\theta
') \eee
is the {scattering kernel}.

The Ward identity \re{Widentitya}, together with the factorisation
\re{dispersionlessf}, imply a {\it non-linear} integral equation for
the dressed expectation values $\e(\th ) \equiv \<
\eb(\th)\>_{_\cyl}$,
 \bee \la{Widentity} \e (\theta ) & = LE(\theta ) +\s\,
 \int_{-\infty}^\infty {d\theta '\over 2\pi} K(\theta -\theta ') \log
 (1- \s\, e^{ - \e (\theta ')}) \, , \eee
 which  is identical  with the TBA equation for the   {\it pseudoenergy}.
 
In the second Ward identity, eq.  \re{Widentitya2}, the expectation
value contains tree Feynman graphs as well as Feynman graphs with one
cycle.  The graphs with one cycle produced by the first and by the
second term cancel and the Ward identity boils down to a {\it linear}
integral equation for the dressed scattering phase $\phi(t) \equiv \<
\phib(t)\> _{\cyl}$,
\bee \la{Wardphi} \p \phi(\th) & = R\p p(\theta ) +
\int_{-\infty}^\infty {d\theta '\over 2\pi} K(\theta -\theta ') { \p
\phi(\th') \over e^{ \e (\theta ')}- \s } \, , \eee
the meaning of which within the TBA approach will be clarified at the
end of this subsection.

From the point of view of Feynman graph expansion of the effective
field theory, summarised in appendix \ref{sec:Feynman}, the non-linear
equation for the pseudoenergy sums up the tree Feynman graphs studied
in \cite{Kostov:2018ckg,Kostov:2018dmi}.  In this respect the
effective field theory is of mean-field type.  Indeed, since the
interaction potential in eq.  \re{partfGENbisTBA} is linear in
$\phib$, the Feynman graph expansion for the free energy stops at one
loop.  Since the quadratic forms for the gaussian fluctuations of
bosons and fermions are the same, the total one-loop contribution
vanishes and only tree Feynman graphs survive.\footnote{This
cancellation takes place only if periodic boundary conditions are
used, i.e. if the cylinder is considered as the large $R$ limit of an
$R\times L$ torus.  If the cylinder is left with open boundaries
supplied with some integrable boundary conditions, then the gaussian
fluctuations of the bosons and the fermions do not cancel but give the
universal part of the boundary entropy.  }
  
As a consequence, the operator $\FF_{\cyl}$ is a dispersionless field
as well, $\< \exp\FF_\cyl \> = \exp\< \FF_\cyl\> $.  This property
implies that the partition function is the exponential of the
expectation value of the free energy operator $\FF_\cyl$.  The
computation of the latter is easy,
 \be \la{freeenergycyl} \CF_\cyl =\< \FF_\cyl \> = -\s R\int
 {d\th\over 2\pi } \log\( 1- \s e^{-\e(\th)}\)\p p(\th).  \ee
The rhs is identical to the expression for the free energy obtained in
the TBA approach.

To complete the correspondence between the effective field theory and
the TBA, let us express, assuming  fermionic TBA statistics, $\s = -1$,
  the dressed expectation values $\phi$ and $\e$  in terms of   the particle and hole densities  $\rho_p$ and
$\rho_h$  as defined in the original paper by Yang and Yang \cite{YY}. 
Obviously   expectation value  $\e$   
corresponds to the pseudoenergy, while the derivative of the phase $\phi$
gives the  density of the available states,
\bee \e &= \log {\rho_h\over \rho_p},\quad \partial_\th \phi =R(
\rho_p+\rho_h)
  .
\eee
Upon this ifentification, the Ward identity for $\e$, eq. \re{Widentity},   becomes  the TBA equation, as mentioned above, 
while   the Ward identity for $\phib$, eq.  \re{Wardphi}, is equivalent
to the linear constraint satisfied by the particle and hole densities
(the Bethe equation in terms of densities)
\be \la{barcritrhorhob} \begin{split}
  \rho_p(\th)  + \rho_h(\th)  &= \p \tilde p(\th) +    \int_ \IR{d\th'\over 2\pi }   K(\th,\th') 
  \rho_p(\th) .
\end{split}\ee

	 \subsection{One-point function in terms of HS fields }

 In infinite volume, all matrix elements of a local operator O can be
 expressed, with the help of the crossing formula, in terms of the
 infinite-volume elementary form factors
 \be \la{elFF} F_n^{\mathcal{O}}(\th_1,...,\th_n) = \< \emptyset |
 \mathcal{O}| \th_1,...,\th_n\>.  \ee
 The elementary form factors for local operators satisfy the Watson
 equations
\be F_n(\th_1, \dots, \th_j, \th_{j+1}, \dots, \th_n)=S(\th_j,
\th_{j+1}) F_n(\th_1, \dots, \th_{j+1}, \th_{j}, \dots, \th_n) \ee and
have kinematical singularities
\be \la{kinsing} F( \th', \th, \th_1, \dots, \th_n) = {i\over i\pi
+\th' -\th} \( 1- \prod_{j=1}^n S(\th, \th_j) \) F_n(\th_1, \dots,
\th_n) + \text{regular}.  \ee
where it is assumed that the infinite volume states are normalised as
$\<\th|\th'\> = 2\pi \d(\th-\th')$.

 The diagonal limit of the form factors for local operators is not
 uniform and depend on the prescription.  The diagonal matrix elements
 are given by the connected form factors $F_{2n}^c(\th_1, \dots,
 \th_n) $ obtained by performing the simultaneous limit $\delta _1,
 \dots, \delta _n\to 0$.  The connected diagonal form factor is
 obtained by retaining the $\delta $-independent part
 \cite{Mussardo-Book},
\be F_{2n}(i\pi+\th_n +i\delta _n, \dots, i\pi+\th_1+ i\delta _1,
\th_1, \dots, \th_n) = F_{2n}^c(\th_1, \dots, \th_n) + \ \delta
\text{-dependent terms}.  \ \ \ \ee

    \begin{figure}[h!]
		 \centering
	 \begin{minipage}[t]{0.50\linewidth}
            \centering
            \includegraphics[width=6.8 cm]{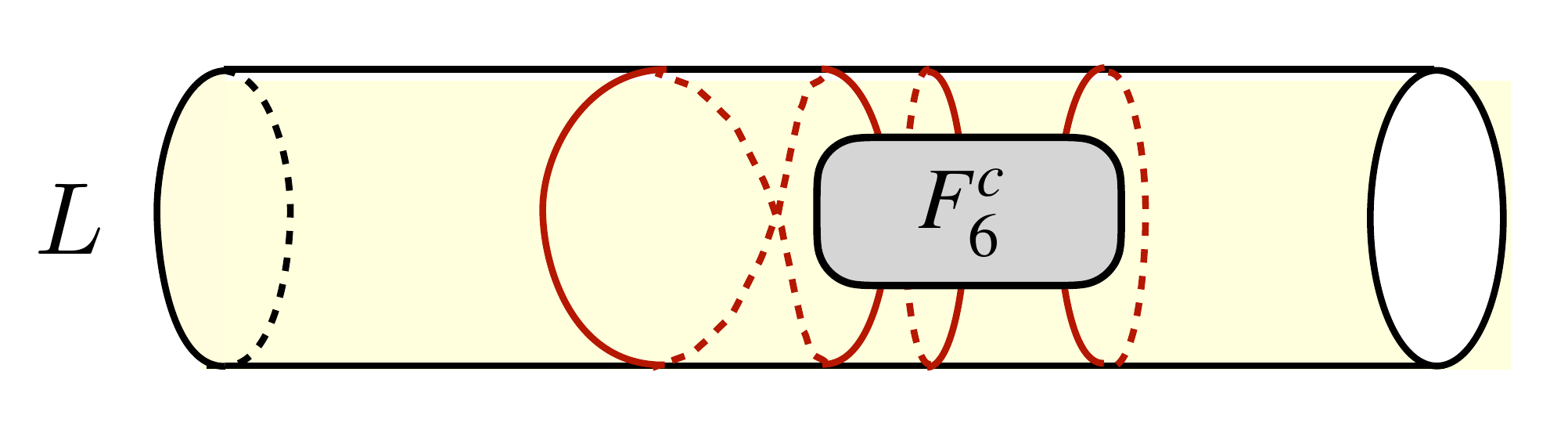}
		  \caption{ \small Three loops with winding numbers $w_1=2$
		  and $w_2=w_3 =1$, attached to a connected form factor
		  $F_6(\th_1,\th_2,\th_3)$.  }
  \label{fig:FF6}
  \end{minipage}
 \end{figure}

Here I will give the field-theoretical equivalent of the derivation
presented in \cite{Kostov:2018ckg}.  In the loop gas representation,
each outgoing particle is identified, after wrapping the cylinder at
least once, with the incoming particle with the same rapidity, as
shown in fig.  \ref{fig:FF6}.  The Boltzmann weight of the loops with
winding number $w_1,...,w_n$ is given by the expectation value
\re{operepsum}.  The sum over the winding numbers is performed without
the symmetry factor $1/|w|$ because there is no more cyclic symmetry.
For the fermionic choice $\s=-1$, one obtains after summing over the
number of particle pairs, the Leclair and Mussardo series
\cite{Leclair:1999ys} for the one-point function of any  local operator,
 \bee \la{LMseries} \< \CO \>_\cyl &=\left\< \sum_{n=1}^\infty {1\over
 n!} \int \prod_{j=1}^n {d\th_j\over 2\pi } \( {1\over
 e^{\eb(\th_j)}+1 } + {e^{\eb(\th_j)} \etab(\th_j) \partial \bxi
 (\th_j)\over (e^{\eb(\th_j)}+1 )^2 } \) F^c_{2n}(\th_1, \dots, \th_n)
 \right\>_\cyl \\
 &= \sum_{n=1}^\infty {1\over n!} \int \prod_{j=1}^n {d\th_j\over 2\pi
 } {1\over e^{\e(\th_j)}+1 }\ F^c_{2n}(\th_1, \dots, \th_n) , \eee
where the cancellation of the bosonic and the fermionic loops is taken
into account.  
This derivation is of course equivalent to 
the one  by the tree
expansion method given  in  \cite{Kostov:2018ckg}.
   As for the higher correlation functions, they can be in principle 
   expressed in loop-gas terms, but the interactions of the loops will be more complicated and it is not clear if and how the formalism developed here can be generalised.

 \section{ Generalised free theories on a torus}
   \la{sec:EQFTtorus}

Before addressing interacting theories, it makes sense to work out the
counting and the statistics of the non-interacting loops on the torus.
In this section, the partition functions of the generalised free
theories, the free neutral massive boson (FB) and the Ising field
theory (IFT), will be derived using the loop-gas formulation of these
theories.  For the FB it is sufficient to compute the path integral of
a loop, while the IFT requires some additional combinatorics.

 \subsection{Path integral for a loop embedded in the  torus}           
  \label{sec:periodicwrappingtor}

\begin{figure}[h!]
  \centering
	 \begin{minipage}[t]{0.85\linewidth}
            \centering
			\includegraphics[width=14.5 cm]{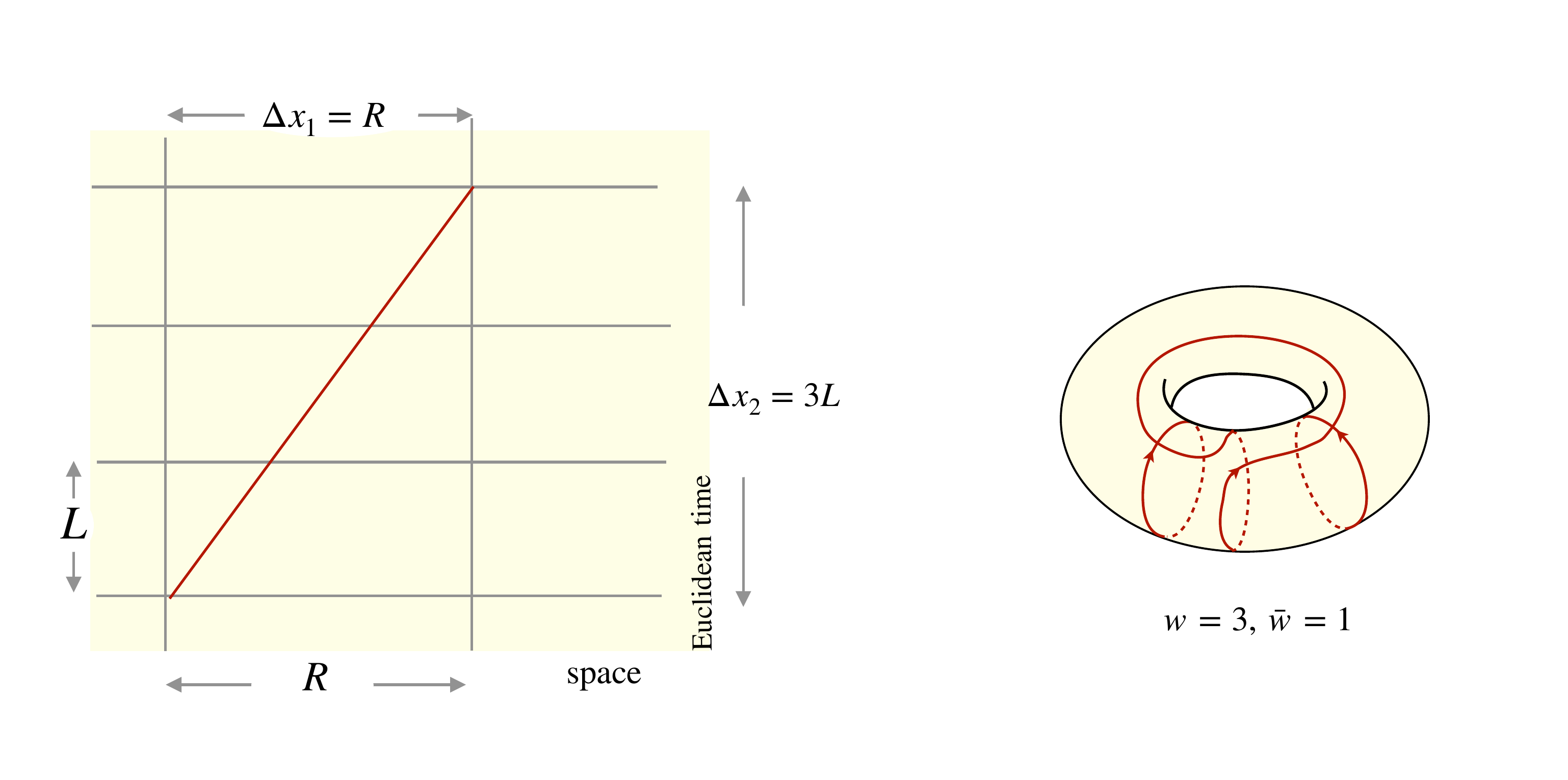}
\vskip -0.5cm \caption{ \small A loop with winding numbers $w , w'$ as
a path integral with a discontinuity $\delta x =\{ w' R, w L\}$ for
$w=3, w'=1$ }
\end{minipage}
  \label{fig:fig3x}
 \end{figure}

Assume that the torus has perpendicular cycles of lengths $L$ and $R$.
The generalisation to the case of a tilted torus is mechanical.  Let
us denote by $[\CF^{(L,R)}]_{w,w'}$ the path integral for a loop in
the topological sector characterised by winding numbers $\{w ,w'\}$.
The explicit expression for this path integral is given by the
integral \re{somewn1} over the constant mode of the 2-momentum, with
discontinuities $\Delta x_1= w' R$ and $ \Delta x_2 = w L$, as
illustrated in fig.  \ref{fig:fig3x},
 \be \la{FDeltaxww} [\CF^{(L,R)}]_{w,w'}= RL\, \mathcal{F} (\delta \vec
 x)\Big|_{\delta x_1= w' R, \delta x_2 = w L}.  \ee
 Let me remind that the  integral  \re{somewn1}  is obtained from the path integral
 for loops in {\it infinite} volume.     
If at least one of the winding numbers is non-zero, in the rhs of
\re{FDeltaxww} given by the double integral \re{somewn1}, one of the
integrations can be performed by residues, thus bringing the remaining
integral on the mass shell.

$\bullet$ If $\delta x_2\ne 0$, the integral with respect to $k_2$ can
be taken by residues and the integrand takes the form of the wave
function of a on-mass-shell particle with momentum $k_1$, propagating
in the {\it direct }channel, analytically continued to imaginary time
$t= - i \delta x_2$,
\bee \la{rlpathintegral-bis-tordirect} \mathcal{F} (\delta \vec x) & =
{\textstyle{1\over 2}}  \, {1 \over |\delta x_2|} \
\int_\mathbb{R} {d p(\th)\over 2\pi} \ e^{ - E(\th)| \delta x_2|+ i
p(\th) \delta x_1 } \qquad (\delta x_2\ne 0) .  \eee

$\bullet$ If $\delta x_1\ne 0$, one can integrate with respect to
$k_1$ by residues, with the result that the integrand takes the form
of the wave function of on-mass-shell particle propagating in the {\it
cross} channel, analytically continued to imaginary time $t= - i
\delta x_1$,
\bee \la{rlpathintegral-bis-tor} \mathcal{F} (\delta  \vec x) & =
{\textstyle{1\over 2}} \,   {1 \over |\delta  x_1|} \int_\mathbb{R}
{d p(\th)\over 2\pi} \ e^{- E(\th) | \delta  x_1| + i p(\th)\ \delta 
x_2 } \qquad (\delta  x_1\ne 0) .  \eee
The two wave functions are related by a double Wick rotation
exchanging the energy and the momentum, known as {\it mirror}
transformation,
 \be \la{mirror1} E\to - i p, \ p\to i E \ee
  or in terms of the rapidity defined in \re{rlpathintegral-bis},
 \be
 \la{mirrortr}
 \theta\ \to\ \ i\pi/2 - \theta.
\ee
The particles propagating in the direct and in the mirror channels are
sometimes referred to a `physical' and `mirror' particles
respectively.
 The mirror transformation \re{mirrortr}  
 relates a  physical particle with winding numbers $w$ and $w'$
 and a mirror particle with 
 winding numbers $w'$ and $w$.

\bigskip

\bigskip

\noindent

Thus the path integral $[\CF^{(L,R)}]_{w,w'}$ admits the following two
integral representations,
\begin{align}
   \la{defgfGFT1} \mathcal{F} _ {w, w' } & \defeq { \hf } R \int
   _{\mathbb{R} } {dp(\th )\over 2\pi} {1\over |w|}\ e^{ -| w| L E(\th
   ) + i w' R p(\th) } \qquad ( w\ne 0), \\
 \la{defgfGFT2} \tilde \CF _{w',w} &\defeq { \hf }L \int _{\mathbb{R}
 } {dp(\th )\over 2\pi} {1\over |w'|}\ e^{ -| w'| R E(\th ) +i w
 Lp(\th)} \qquad ( w'\ne 0) .
    \end{align}
If $w'=0$ or $w=0$, only one of the two representations makes sense.
When both winding numbers are non-zero, there is a choice between
\re{defgfGFT1} and \re{defgfGFT2}.  By construction,
   \be \la{F=Fdual} \CF _{w,w'}= \tilde \CF _{w',w} \qquad (w, w'\ne
   0).  \ee
 This identity can be also proved directly   by shifting the contour of
integration for $ \CF_{w ,w' } $  by $i\pi/2$ 
 and integrating by parts.
  Using the freedom of choice between $\CF_{w ,w' }$ and
  $\tilde \CF _{w',w} $ when bith winding numbers are non-zero, 
  one can write different integral representations
 for the free energy of the loop gas.

 \subsection{ Free massive boson }
  \label{sec:periodicwrappingfb}

The QFT of a massive bosonic field on the $L\times R$ torus is defined
by the Euclidean path integral \re{partffreeb} with doubly periodic
boundary conditions $ \vp(x_1+R,x_2 ) = \vp(x_1, x_2) $ and $
\vp(x_1+L,x_2 ) = \vp(x_1, x_2).  $
The loops being non-interacting, the free energy is equal to the path
integral for a single loop.  The path integral includes a sum over the
topological sectors with winding numbers $w$ and $w'$,

\bee \la{rlpathintega} \mathcal{F}_\FB^{(L,R) } & =\sum_{ w , w'
\in\IZ }[\mathcal{F} ^{(R,L)} ]_{ w,w'} .  \eee
The following three choices for the sum in \re{rlpathintega} lead to
three different integral representations of the free energy of the
loop gas,
\begin{align}
\la{wwsum1} \FFB ^{(L,R)}_\FB & = \sum_{ w \ne 0 } \FFB _{ w,0} +
\sum_{w'\ne 0 } \sum_{w\in\IZ}\FFBT _{w',w } \\
\la{wwsum2} & =\sum_{ w' \ne 0 } \tilde{\mathcal{F}} _{ w',0}
+\sum_{w'\in\IZ} \sum_{ w\ne 0} \FFB _{w,w' } \\
 \la{wwsum3} & = \sum_{w' \ne 0} \FFBT _{ w',0}+ \sum_{w\ne 0} \FFB
 _{w,0} + \hf \sum_{w,w'\ne 0} \( \mathcal{F} _{w,w'} + \FFBT _{w',w }
 \) .\end{align}

Let us start with the choice \re{wwsum1}.  The first sum has been
calculated in the previous section.  In the second double sum, the sum
over $w' $ gives the periodic delta function $\d^{(2\pi)}\!\left[ L \,
p(\th) \right] $.  By substituting the holomorphic representation of
the delta function, the second piece is written as a contour integral,
\bee \la{sumFrl} 2\sum_{w' =1}^\infty \sum_{ w\in\IZ}
\tilde\CF _{w' , w} &=- \oint _{ \CC_{\IR} } \log\left( 1-
e^{-R E(\th )}\right) \ {d\log \(1- e^{ i L p(\th) }\) \over 2\pi i}\,
, \eee
 where $\CC_\IR$ is a contour enclosing the real axis,
  \be \la{defCCR} \mathcal{C} _\mathbb{R} = \{ \overset{-\!-
  \!\rightarrow}{\mathbb{R} -i0}\}\cup \{ \overset{ \leftarrow\!-\!-
  }{\mathbb{R} +i0}\}.  \ee
Combining the two pieces, one finds for the free energy of the loop
gas the following integral expression,
   \bee \la{FrenFBtor} \mathcal{F}^{(L,R)}_\FB & = - \int _\mathbb{R}
   { R d p(\th)\over 2\pi} \ \log\left( 1- e^{- L E(\th)}\right) -
   \oint _{ \CC_{\IR} }  \log\left( 1- e^{-R E(\th )}\right) {d\log \left( 1-
   e^{ iL p(\th) }\right) \over 2\pi i}.  \eee
 As the integrand of the second term in \re{FrenFBtor} falls
 exponentially at $\th\to \pm\infty$, the contour integral can be done
 by residues, and one obtains the partition function of the massive
 boson in the form given in \cite{Klassen:1990dx},
 \be \la{partftortr} \mathcal{Z} _\FB ^{(L,R)}= { e^{{\pi \over 6}
 {R\over L} c _ 0(mL) }\over \prod_{n\in\mathbb{Z} } (1- e^{- R
 E_n})}, \quad E_n = \sqrt{p_n^2 + m^2}, \quad p_n = {2\pi n\over L}.
 \ee
Thus the two integrals in \re{FrenFBtor} have quite different meaning.
The first one gives the ground state energy in the cross channel,
while the second one takes into account all excited states in the
direct channel.

Now let us consider the other two choices, starting with \re{wwsum2}.
Proceeding as above, one arrives at the dual integral representation,
obtained from \re{FrenFBtor} by exchanging $L$ and $R$,
   \bee \la{FrenFBtordual} \mathcal{F}^{(R,L)}_\FB & = - \int
   _\mathbb{R} { L d p(\th)\over 2\pi} \ \log\left( 1- e^{- R
   E(\th)}\right) - \oint _{ \CC_{\IR} } \log\left( 1- e^{-L E(\th
   )}\right) {d\log \left( 1- e^{ iR p(\th) }\right) \over 2\pi i}.
   \eee
 In the dual representation, the first term is proportional to the
 ground-state energy in the direct channel while the second term takes
 into account the contributions of the excited states in the cross
 channel.

Finally the third choice, eq.  \re{wwsum3}, gives a self-dual integral
representation which can be written, using the mirror transformation
\re{mirrortr}, only in terms of the energy,
 \bee \la{Fsimbos} \mathcal{F} ^{(L,R)}_\FB &= - \int\limits _{
 \mathbb{R} -i0 } \log\left( 1- e^{-L E(\th )}\right) { d\log \left(
 1- e^{ - R E(i\pi/2-\th) }\right) \over 2\pi i} +\{L\leftrightarrow
 R\} .  \eee
 The equivalence of the three integral representations can be also
 established directly by deforming the integration contours and then
 integrating by parts. (There are no surface terms because the integrand 
 decays exponentially at infinity.)

			\subsection{ Ising Field Theory }
  \label{sec:pIFT}

The IFT is a theory of a bosonic particle describing the order
variable of the Ising model in the scaling limit.  Its infinite-volume
thermodynamics is that of a free Majorana fermion, since both theories
describe a free fermionic particle from the TBA perspective.  On the
torus, however, the IFT is not equivalent to the QFT of a Majorana
fermion, as pointed out in \cite{Klassen:1990dx}, although a precise
map exists.

The IFT partition function was first computed by Ferdinand and Fischer
\cite{PhysRev.185.832} using the lattice formulation.  A
field-theoretical derivation (both for the IFT and for the FB) was
given by Itzykson and Saleur \cite{Saleur1987}, who computed the
zeta-function regularised determinant of the Laplace operator on the
torus.  Afterwards Klassen and Meltzer \cite{Klassen:1990dx}
expressed, using TBA-related arguments, the results of
\cite{Saleur1987} in a compact and elegant form.

 It has been known for a long time that the Ising model can be
 reformulated as an ensemble of loops.  Natalya Vdovichenko
 \cite{vdovichenko1965calculation} found out that the sum over the
 Ising clusters arising in character expansion of the partition
 function in the disordered phase ($T>T_c$) can be represented as a
 sum over loops on the lattice with minus signs associated with the
 intersections.  The same reformulation can be done in the ordered
 phase ($T<T_c$), where the clusters are composed of domain walls
 separating domains of up and down Ising spins.  The Boltzmann weights
 of the loops in the ordered and in the disordered phases are related
 by the Kramers-Wannier duality \cite{PhysRev.60.252}.  Vdovichenko's
 method was generalised to the torus in \cite{Morita_1986}, see also
 \cite{WOLFF2020115061}.  Again, in the Boltzmann weight of a loop
 configuration each crossing contributes a factor $(-1 )$.
 Symbolically, the loop gas partition function in the two phases reads
 (only the sign factors are noted)
 \bee
 \la{partfFMF}
 \CZ^{(L,R)}_{\text{IFT}} 
&= \sum_{\text{ loops}} 
  (-1)^{\#\text{crossings}} \ \quad \ \ \qquad\qquad\qquad\qquad\qquad
  (T> T_c) \\
 &= \sum_{\text{ loops}} {\textstyle { 1+(-1)^{\wtot}\over 2}{1+
 (-1)^{\wtot' }\over 2}}
  (-1)^{\#\text{crossings}} \ \ \ \qquad \ \ (T< T_c)
  .
 \eee
 In the second line, $w_\tot $ and $w_\tot ' $ are the total winding
 numbers for the two periods of the torus.  The extra factor in the
 ordered low-temperature phase projects to the loop configurations
 with even total winding numbers in both periods, because periodic
 boundary conditions for the Ising spins are compatible only with even
 number of domain walls.

\bigskip

\noindent $\bullet$ {\it Derivation of the partition function from the
loop gas}

\smallskip\smallskip

The derivation here is along the lines of the lattice derivation given
in \cite{Morita_1986, WOLFF2020115061}, which simplifies considerably
in the continuum limit.  To begin with, let us notice that for a loop
with winding numbers $\{w, w' \}$, the number of times it intersects
itself is \cite{diFrancesco1987}
\be \la{selfint} \# \text{crossings} = w\wedge w' -1 , \qquad w\wedge
w'=\text{the greatest common divisor of $w$ and $w' $}.  \ee
 By definition $w\wedge 0=w$.  Since only the parity of the number of
 intersections is important, it is convenient to use the identity (for
 a single loop)
 \be \la{selfint1} (-1)^{ \# \text{crossings}} = (-1)^{ w w'-w-w'+1} .
 \ee
  From here one infers with little effort that for a configuration of
  $N$ loops with winding numbers $ w_1,..  , w_N$ around the $L$-cycle
  and $w'_1,..., w'_N$ around the $R$-cycle, the sign factor is
 \bee \la{nucross} (-1)^{ \# \text{crossings}}\ & = (-1)^{ {w_\tot
 w'_\tot } -{w_\tot } -{ w'_\tot } +N } \eee
 where $w_\tot =w_1+...+w_N $ and $ w'_\tot = w'_1+...+w'_ { N} $.
 With the help of the identity \re{nucross}, the sign factors in the
 partition sum \re{partfFMF} can expressed solely in terms of the
 total number of loops and the total winding numbers,
 \bee \la{loopgasisingb} \CZ^{(L,R)}_{\text{IFT}} &=
 \sum_{\text{loops}} (-1)^{w_\tot +w_\tot ' +w_\tot w_\tot ' } \
 (-1)^{\#\text{ loops}} \ \ \ \ \ \ \quad \qquad (T>
 T_c) , \\
	&= \sum_{\text{loops}} {\textstyle { 1+(-1)^{\wtot} \over 2}{1+
	(-1)^{\wtot ' }\over 2}} 
	(-1)^{\#\text{ loops}} \ \ \ \quad\qquad\quad\ \ (T< T_c) .  \eee

 Unlike the loop gas for the free boson, the sum over loops in
 \re{loopgasisingb} does not exponentiate because the loops interact
 through the sign factor $(-1)^{w_\tot w_\tot ' }$.  Instead, the sum
 over loops splits into four blocks
 \be \la{defDD} D_{ \a\b}= \sum_{\text{ loops}} e^{ 2\pi i\a w_\tot +
 2\pi i\b w_\tot ' } \ (-1)^{\#\text{ loops}} \qquad ( \a,\b =0,\hf),
 \ee
with the sign factor being constant in each block.  One verifies
immediately, comparing the weights in \re{loopgasisingb} and
\re{defDD}, that
  \bee \la{IFT-F} \CZ ^{(L,R)}_{\text{IFT}} \ =\ \
   \begin{cases}
	 \hf \(D_{{1\over 2} {1\over 2}}+ D_{{1\over 2} 0}+ D_{0 {1\over
	 2} }- D_{0 0}\) & \text{if } \ T>T_c ,\\
	 \hf \(D_{{1\over 2} {1\over 2} }+ D_{{1\over 2} 0}+ D_{0{1\over
	 2} }+ D_{00}\) & \text{if } \ T<T_c.
\end{cases}
\eee
The expansion \re{IFT-F} of the IFT partition function in four blocks
is in accord with \cite{Saleur1987} for the ordered phase and with
\cite{Klassen:1990dx} for the disordered phase.  The different signs
of the last term in the two phases have simple explanation.  The block
$D_{00}$ vanishes at the conformal point $T=T_c$ and is linear in $
T_c-T $ in the scaling regime.  Since the partition function at finite
volume is analytic in $T$, it must have the same analytic form below
and above the critical point.  Thus the different signs come from the
non-analyticity of the mass $m\sim |T-T_c|$.

Let us now compute the four blocks from the loop gas.  Since the loops
in each block do not interact, the free energy is equal to the path
integral for a single loop with sign factors depending on the winding
numbers,
  \bee \la{FrenCINTLL-IFT} \log D_{\a\b} &= - \sum_{ w,w' \in\IZ }
  e^{2\pi i \a w + 2\pi i \b w' }\ [\mathcal{F} ^{(R,L)} ]_{ w,w'} .
  \eee
  As in the case of the free boson, there are two natural ways to
  evaluate the sum, resulting in two integral representations of the
  rhs analogous to \re{FrenFBtor} and  \re{FrenFBtordual}
  for the free boson.  For example, the first one reads
 \bee \la{FrenCINTferm} \log D_{\a\b} &= R \int_{\mathbb{R} } {d p(
 \th )\over 2\pi }\log\left( 1- e^{2\pi i \a} e^{-L E(\th)}\right) \\
		&+ \oint  _{ \CC_{\IR} }  \log\left( 1- e^{2\pi i \b}
		e^{-RE(\th )}\right) { d\log \(1- e^{2\pi i \a} e^{i L p(\th)
		}\)\over 2\pi i} .  \eee
In \cite{Saleur1987,Klassen:1990dx}, the blocks $D_{\\a\b} $ were
computed as the partition functions of a Majorana fermion with various
boundary conditions.  The periodic (Ramond) and the anti-periodic
(Neveu-Schwarz) boundary conditions corresponds respectively to $\a=0$
and $\a={\textstyle{1/2}} $.  To connect with
\cite{Saleur1987,Klassen:1990dx}, one should perform the contour
integral in the second line of \re{FrenCINTferm} by residues, which
gives
   \bee \la{DiscrOscTors} D_{\alpha ,\beta} & = e^{ - { \pi R\over 6 L
   } \ e^{2\pi i \a} c_\a(mL) } \prod_{n\in\mathbb{Z} +\a} \left( 1-
   e^{2\pi i\b }\ e^{- R E_n}\right) \qquad (\a,\b = 0, \hf).  \eee
The scaling function
	 \bee \la{defEEb} c_\a(mL) = -e^{-2\pi i\a}\ {6L\over
	 \pi}\int_{\IR} {d p( \th )\over 2\pi }\log\left( 1- e^{2\pi i \a}
	 e^{-L E(\th)}\right) \eee
is the effective central charge in the cross channel for boundary
condition $\a$, and the product in \re{DiscrOscTors} takes into
account the excited states in the direct channel.  In the UV limit
$m\to 0$, the effective central charge tends to its ultraviolet value
  \bee c_\a (mL) & \underset{ {m=0} } \to\ =1- \a \qquad\quad  (\a =0,
  \textstyle{1\over 2}).  \eee
In the deep IR limit, $m\to\infty$, there are no degrees of freedom
left and the effective central charge vanishes exponentially.  The
partition function \re{IFT-F} in this limit counts the number of the
ground states, 1 in the disordered phase and 2 in the ordered phase.
 
\section{ Integrable QFT on a torus }
\la{sec:WindIntTheor}

\subsection{Mapping to a loop gas }
\la{section:interloopgas}
 
 After getting the counting and the statistics of the loops straight,
one can proceed by switching on the dynamics, as it was done for the
theory on a cylinder.  Let us consider a
 non-trivial scattering theory compactified on a rectangular $L\times
 R$ torus.  The claim is that the partition function of an interacting
 QFT of the type considered in section \ref{section:loopgascyl} can be
 computed as the grand partition function of a gas of loops immersed
 in the torus, with two-body interactions associated with the
 intersections.  Each intersection is counted with a scattering factor
 which depends on the rapidities of the two intersecting segments of
 loops.

 As was emphasised in section \ref{sec:periodicwrapping}, the path
 integral of a winding loop can be expressed in terms of the wave
 functions either of `physical' on-shell particles propagating in the
 direct channel, eq.  \re{defgfGFT1}, or in terms of the wave
 functions of `mirror` on-shell particles propagating in the cross
 channel, eq.  \re{defgfGFT2}.  In what follows, the terms `physical
 particle' and `mirror particle' applied to a loop will signify which
 of the two choices is taken.

It became clear from the analysis of the generalised free theories
that the loop-gas formulation necessarily involves both physical and
mirror particles.  When two particles with rapidities $\th_1$ and
$\th_2$ have the same kinematics (physical-physical or mirror-mirror),
their crossings are weighted by scattering matrix $S(\th_1-\th_2)$.
When the two particles have different kinematics (physical-mirror or
mirror-physical), their crossings are weighted by the scattering
factor with one of the arguments mirror-transformed by \re{mirrortr}.
Then the weight is a function of the {\it sum} of the two rapidities,
$W(\th_1+\th_2)$, defined as
  \be
  \la{defW}
  \W(\th)\defeq S(\th  - i {\textstyle {\pi\over 2}}).
  \ee
  The function $W$ is a real analytic function in the whole
  $\th$-plane except for a periodic array of poles on the imaginary
  axis.  It has the symmetries
 \be \la{symplase} \W (\th ) = \W ( -\th ) = \W ( \th +i\pi ) ^{-1} =
 \W (\th ^*)^* , \ee
 which follow from the unitarity, crossing symmetry and the real
 analyticity of the S-matrix, eq.  \re{proprsS}.

Let us give a precise formulation of the statistical ensemble of
interacting loops.  Take the most general configuration with $N$
particles in the direct channel with rapidities $\th_j$ and winding
numbers $ \{ w_j, w' _j\} $, and $\tilde N$ particles in the cross
channel with rapidities $\tilde \th_k$ and winding numbers $ \{ \tilde
w _{k}, \tilde w'_{k} \} $.  The Boltzmann weight of this
configuration is, according to \re{defgfGFT1} and \re{defgfGFT2} ,
\bee \la{ZNNone} \CZ^{{N,\tilde N}}_{[{ \{\th_j; w_j, w'_j\} ,
\{\tilde \th_k; \tw_k, \tw'_k\}} ] } &= \prod_{j=1}^N { e^{-L |w_j|
E(\th_j) +i R w' _j p(\th_j)}\over 2 |w_j|} \prod_{k=1}^{\tilde N}
{e^{-R| \tw _k| E(\tilde \th_k) +i L\tw '_k p(\tilde \th_k)} \over
2|\tw_k|} \\
  & \times \prod_{j=1}^N \s^{w_j+w'_j +1} \prod_{k=1}^{\tilde N}\s^{\tw_k +\tw'_k+1}  
  \\
&\times
\prod_{i,j   =1}^N  
S(\th_{i}-\tilde \th _{j})^{ |w' _{i}|\,  w_{j} - |w'_{ j}|\,  w_{i}       } 
\prod_{ k,l =1}^{\tilde N } 
S(\th_{k}-\tilde \th _{l})^{ |\tw' _{k}|\,  \tw_{l} - |\tw'_{ l}|\,  \tw_{k}       } 
\\
&\times 
\prod_{j=1}^N\prod_{k=1}^{\tilde N}  W(\th_j+\tilde \th_k)^{-|w_j ||\tw_k| + w'  _j \tw ' _k}
S(\th_j-\tilde \th _k)^{ w' _j   | \tw_k|-|w_j| \tw _k'      } 
 .
 \eee

Without scattering, the integration measure would be 
a product of the one-particle integration measures,
\be
\la{measurefortheloops}
 \prod_{j=1}^N {Rdp(\th_j)\over 2\pi} \prod_{k=1}^{\tilde N} {L
dp(\tilde \th_k)\over 2\pi} = \prod_{j=1}^N d\phi _j
\prod_{k=1}^{\tilde N} d\tilde\phi_k\, , \ee
where the rhs is expressed through the phase shifts of the wave
functions of the physical and the mirror particles.  
If both $L$ and
$R$ are large, the measure is again the flat measure for the phase
shifts which now have contribution from the scattering\footnote{Of
course, after the analytic continuation, the phase shifts can become
complex.}
  \bee
  \la{phiphitilde}
   \phi_j & =Rp(\th_j)- i \sum_{j'=1}^N |w_{j '} | \log
  S(\th_j-\th_{j'}) - i \sum_{k=1}^{ \tilde N} \tw ' _k \log W(
  \th_j+\tilde \th_k ), \quad (j=1,..., N)\\
  \tilde \phi_k  & = Lp(\tilde \th_k) - i
  \sum_{j=1}^N w' _{j } \log W(\tilde \th_k+\th_{j}) - i \sum_{k'=1}^{
  \tilde N } |\tilde w_{k'}| \log S( \tilde \th_k-\tilde \th_{k'} )
  \quad (k=1,..., \tilde N).
  \eee
 Since all winding effects are already taken 
 into account,  it is plausible that the expression  \re{measurefortheloops}
 holds also for finite $R$ and $L$, although I do not know if a rigorous proof  can be constructed at all.  With  no proof available,   
 the expression \re{measurefortheloops}  should be considered as a 
  {\it definition}  of  the measure   for the loop gas.  
 
  With the measure thus defined,   the contribution to the partition 
function of the collection of loops with this particular topology is
 given by the integral of the $N+\tilde N$ rapidities 
 \bee \la{NNamplit} \CZ^{{N,\tilde N}}_{[{ \{ w_j, w'_j\} , \{ \tw_k,
 \tw'_k\}} ] } &= \ \int\limits_{\IR^{ N+\tilde N }} \CZ^{{N,\tilde
 N}}_{[{ \{\th_j; w_j, w'_j\} , \{\tilde \th_k; \tw_k, \tw'_k\}} ] } \
 \prod_{j=1}^N { d\phi_ j\over 2\pi}\prod_{k=1}^{\tilde N} {d\tilde
 \phi_k\over 2\pi},  \eee
 with $\phi_j$ and $\tilde \phi_k$ given by \re{phiphitilde}.
  The integration measure contains a Jacobian for the change of variables from
  $\{\phi, \tilde \phi\}$ to $\{\th,\tilde \th\}$:
 \bee \la{GaudinDT} \prod_{j =1}^N {d \phi_j } \prod_{k=1}^{\tilde
 N}{d \tilde \phi_k } = \prod_{j =1}^N {d\th_j } \prod_{k=1}^{\tilde
 N}{d\tilde \th_k } \ \det \left(\begin{array}{cc} {\p\phi_j/\p
 \th_{j'} }& {\p\phi_j/\p \tilde \th_k} \\ {\p\tilde \phi_k / \p\th_j}
 & {\p\tilde \phi_k/ \p \tilde \th_{k'}}\end{array}\right) .  \eee

The loop amplitudes of the type \re{ZNNone} split into equivalence
classes related by mirror transformations with respect to part of the
rapidities.  The grand canonical partition functions must contain only
one representative of each class.  A possible choice for the sum,
corresponding to the choice \re{wwsum1} for the free  boson, is
\bee \la{grandPF} \CZ^{(L,R) }_\tor&=\sum _{N,\tilde N=0}^\infty
\sum_{\{w_j\ne 0 \}} \sum _{\{\tw_k \ne 0, \tw'_k\}} \ {\CZ^{N,\tilde
N}_{[{ \{w_j, 0\} , \{\tw_k, \tw'_k\}} ] }\over N!\ \tilde N!}
\\
&\underset{S=1}
\to \exp\(    \sum_{ w \ne 0 } \FFB _{ w,0} +
\sum_{\tw'\ne 0 }\sum_{\tw\in\IZ} \FFBT _{w',w }\).  \eee

\subsection{ The partition function in terms of HS fields}
\la{section:auxfildstor}
 
The effective QFT for the torus can be constructed as a generalisation
of the operator representation \re{operccyl} for the cylinder.  To
achieve a symmetric description of the direct and the cross channels,
it is convenient to introduce special notations for the mirror images
of fields $\eb$ and $\phib$ defined in section \re{gaussrepcyl},
   \be \la{dualpairop} \tphib(\th) \equiv i \eb(i\pi/2 - \th), \qquad
   \teb(\th) \equiv - i \phib(i\pi/2-\th).  \ee
 The HS  fields are designed to generate the scattering factors
 in \re{ZNNone}.  For that they should have the following two-point
 functions,
	\bee \la{defgfGFT11} & \langle \phib ( \theta ) \eb (\theta ')
	\rangle _c = \langle \tphib ( \theta ) \teb (\theta ') \rangle _c
	= i \log S( \theta -\theta ')\, , \\
	 & \langle \eb (\th) \tilde\eb ( \th ' ) \rangle_c = \langle \phib
	 (\th) \tphib ( \th ' ) \rangle_c=- \log \W ( \th + \th ') \, , \\
	& \< \eb(\th) \eb(\th') \>_c =\< \phib(\th) \phib(\th') \>_c = \<
	\teb(\th) \teb(\th') \>_c =\< \tphib(\th) \tphib(\th') \> _c =\<
	\eb(\th) \tphib(\th') \>_c= \< \teb(\th) \phib(\th') \> _c=0.
	\eee
Of course, the propagators involving $\tilde \eb$ and $\tilde \phib$
follow from the definition \re{dualpairop}.  The asymptotic energies
and momenta are introduced as classical values of the HS 
fields, 
\be \la{expvalaux} \langle \eb (\theta)\rangle = L E(\theta),\ \ \
\langle \phib (\theta)\rangle = R p(\theta) ,\quad \langle \teb
(\theta)\rangle = R E(\theta),\ \ \ \langle \tphib (\theta)\rangle = L
p(\theta) \, .  \ee
From the properties of the two-point correlators and the classical
values it follows that the HS  fields are real analytic and
anti-periodic with respect to $\th\to\th+i\pi$.
This will be used later to construct oscillator  representation
of the HS  fields associated with their expansions  in the odd powers of $e^\th$.

The path integrals for loops with winding numbers $w,w'$, evaluated in
physical and in mirror kinematics, eqs.  \re{defgfGFT1} and
\re{defgfGFT2}, are now replaced respectively by the operators
 \begin{align} \la{boltzmweightsbIFT1} \FF_{w ,w' } & = \hf \, \s^{w+w'-1}
  \int_{\mathbb{R} } {d\phib(\th )\over 2\pi} {1\over |w|}\ e^{ -| w|
  \eb(\th ) + i w' \phib(\th)} \qquad (w\ne 0) \\
	\la{boltzmweightsbIFT2} \tilde \FF_{w' ,w } & = \hf \, \s^{w+w'-1}
	\int_{\mathbb{R} } {d\tphib(\th )\over 2\pi} {1\over |w'|}\ e^{ -|
	w'| \teb(\th ) + i w \tphib(\th)} \qquad (w'\ne 0).
  \end{align}
  The  operators  $\FF_{w ,w' } $ and  $\tilde \FF_{w' ,w } $   
  satisfy the operator analogue of the duality relation     \re{F=Fdual},
  \be \la{FFFFT} \FF_{w, w'}=\FFT_{w', w}\qquad (w, w'\ne 0),
   \ee
 which can be proved as  \re{F=Fdual}   by shifting the contour of
integration on the lhs by $i\pi/2$ and integrating by parts.

   Now  the integral \re{NNamplit} takes the form of  the  following expectation
 value
  \bee \la{ZNNone1} \CZ_{ \{w_j, w'_j\}, \{\tw_k, \tw'_k\}} &=\left\<
  \prod_{j =1}^N \FF_{w_j, w'_j} \prod_{j =1}^N \FFT_{\tilde w_j,
  \tilde w'_j} \right\> . \eee
  To express the operator differentials in terms of derivatives, one
  can proceed as in section \ref{gaussrepcyl} by introducing
  Faddeev-Popov ghosts.  Now we have two pairs of fermions $\{\bpsi,
  \etab\}$ with the same two-point functions as the bosons with
  correlation functions,
   \bee \la{defetapsitor} \langle \etab (\th ) \bxi (\th')\rangle &=
   \langle \tetab (\th ) \tbpsi (\th')\rangle = i \log S( \th-\th'),
   \\
	\langle \etab (\th ) \tbpsi (\th')\rangle &= \langle \tetab (\th )
	\bxi (\th')\rangle = i \log \W( \th+\th') .  \eee
The fermions $ \tetab $ and $\tbpsi $ are mirror images of $ \bpsi$
and $ \tbpsi$,
   \be \la{dualpairferm} \tetab(\th) \equiv \bxi (i\pi/2 - \th),
   \qquad \tbpsi(\th) \equiv \etab (i\pi/2-\th).  \ee

The fermionic representation of the Jacobian on the rhs of
\re{GaudinDT} is a straightforward generalisation of eq.
\re{operepsum}.  If the product of differentials in the expectation
value in \re{ZNNone1} is replaced as
	 \bee \la{FrenergyOp} \prod_{j=1}^{ N}{d\ \phib_j \over 2\pi }
	 \prod_{k=1}^{\tilde N}{d\tilde \phib_k \over 2\pi } & \to\
	 \prod_{j=1}^{ N}{d\ \th_j \over 2\pi } \prod_{k=1}^{\tilde
	 N}{d\tilde \th'_k \over 2\pi } \\
	  &\times \prod_{j =1 }^N { \p \phib (\th_j) - |w_j| \, \etab (
	  \th_j) \p \bxi(\th_j) \over 2\pi } \ {\p \tphib (\th'_k) - | w'
	  _k| \, \tetab( \th'_k) \p \tbxi (\th'_k) \over 2\pi} , \eee
then the expectation value of the rhs gives the Jacobian on the rhs of
\re{GaudinDT}.  The corresponding expressions for the operator loop
amplitudes \re{boltzmweightsbIFT1} and \re{boltzmweightsbIFT2} are
	  \bee \la{boltzmweightsbIFTff} \FF_{w ,w' } & = \hf \,
	  \s^{w+w'-1} \int_{\mathbb{R} } {d\th\over 2\pi} \ e^{ -| w| \eb
	  + i w' \phib } \( { \p_\th \phib \over |w|} -\etab \p _\th \bxi
	  \) \qquad (w\ne 0) \\
   \tilde \FF_{w ,w' } & = \hf \, \s^{w+w'-1} \int_{\mathbb{R} }
   {d\th\over 2\pi} \ e^{ -| w| \teb + i w' \tphib } \( { \p_\th
   \tphib \over |w|} - \tetab \p_\th \tbxi \) \qquad (w\ne 0).  \eee

 The sum over the winding numbers can be performed explicitly as in
 the case of the free theory.  The partition function is equal to the
 expectation value
   \be \la{partfGENstar} \mathcal{Z} _\tor^{(L,R)}= \langle
   e^{\mathbf{F} _\tor}\rangle , \ee
  where the operator $ \FF_\tor $ can be given different integral
  representations depending on the choice physical/mirror for the
  winding loop kinematics.  Performing the sum as in \re{wwsum1}, one
  obtains
 \bee \la{GFTcfirstnew1} 
 \FF_\tor = & -\s \int_{\mathbb{R} } {d \th \over 2\pi } \left[
 \log\left( 1 -\s e^{- \eb }\right) \partial _\th \phib \, + { \etab
 \p _\th \bxi \over 1- \s\, e^{ {\eb } } } \right] \\
 & -\s \oint _{ \CC_{\IR} } {d \th \over 2\pi } {1\over 1- \s
 e^{-i\tphib } } \left[ \log\left( 1 -\s\, e^{- \teb } \right) {
 \partial _\th \tphib } +{ \tetab \partial_\th \tbxi \over 1-\s e^{
 {\teb } } }\right] \, .  \eee
%
 %
%
The dual representation, which corresponds to the sum as in
\re{wwsum2} and is mirror image of the first, reads
%
\bee \la{GFTcfirstnew2} \mathbf{F} _\mathrm{tor} = & -\sigma  \int \limits_\mathbb{R}
{d\theta \over 2\pi} \[ \log\left( 1-\sigma  e^{- \teb(\theta)}\right)
\partial_\theta\tphib + { \tetab \partial_\theta \tbxi \over 1-\sigma  e^{
{\teb } } }
	 \]
	 \\
	 & -\sigma  \oint \limits _{ \CC_{\IR} } {d \theta \over 2\pi } {1\over 1-
	 \sigma  e^{-i\varphi  } } \left[ \log\left( 1 -\sigma \, e^{- \eb } \right) {
	 \partial _\theta \varphi  } +{ \etab \partial_\theta \bxi \over 1-\sigma  e^{
	 {\eb } } }\right] .  \eee

The operator $\FF_\tor$ can be treated as the interaction
potential in a standard quantum field theory. 
 In particular, one can
set up a Feynman diagram expansion of the free energy above the
mean-field given by the TBA equation.
The integral representation \re{GFTcfirstnew1} of the interaction
potential is best suited for calculating the exponential corrections
to the TBA solution when $L$ is finite and $R$ is  
large.  The exponential corrections in this limit come from the second
line in \re{GFTcfirstnew1}.  In the opposite limit where $L$ is large
and $R$ is finite, one can use the dual integral
representation \re{GFTcfirstnew2}.

Let us describe qualitatively the diagram technique, cf. appendix \ref{sec:Feynman}.
It makes sense to expand around the mean-field solution
in order to avoid summing over trees.
The mean-field solution
which is determined by a pair of non-linear integral equations
\bee
\la{Widentityator}
 \e(\th)  & = LE(\theta )
-  \int_{-\infty}^\infty {d\theta '\over 2\pi} K(\theta -\theta ')
 \log \(1+ e^{ - \e (\theta ')}\)   
  \\
&-\oint \limits
	 _{ \CC_{\IR} } 
	  {d \th' \over 2\pi } \log W(\th+\th')
	  {  \partial _\th \tilde \phi(\th')\over   (1+ e^{-i\tilde \phi } ) (  1 +\, e^{- \tilde \e (\th') } ) }
 \,
 \\
  \phib(\th) & = R  p(\theta ) +
i \int_{-\infty}^\infty {d\theta '\over 2\pi} \log S(\theta -\theta ')
 { \p \phi (\th')\over 1+e^{  \e (\theta ')} }  \,  
 \\
&- \oint \limits
	 _{ \CC_{\IR} } 
	  {d \th' \over 2\pi } 
  \log W(\th+\th')   {\p \tilde \e (\th')\over (1+e^{\tilde \e (\th')})(1+e^{-i \tilde\phi(\th')})},
  \\
  \tilde \e(\th) & =- i \phi(i\pi/2 -\th), \quad
  \tilde \phi(\th) =i \e( i\pi/2-\th).
\eee
The difference with the (exact) mean-field equations in the TBA limit,
\re{Widentity} and \re{Wardphi}, is in the last terms, which reflect
the presence of excited stated in the cross channel.  The connected
vacuum Feynman graphs have at least one loop.  The total contribution
of the graphs with only one loop vanishes because of the cancellations
between bosons and fermions.  More generally, the contribution of a
graph containing a bosonic loop which is simply connected to the rest
of the graph, is compensated by a similar graph containing a fermionic
loop.

  \subsection{TBA limit  and excited states}
    
  Let us see how the expectation value \re{partfGENstar} generates the
  sum over all excited states in the cross channel.  Choose fermionic
  TBA statistics, $\s=-1$.  In the limit of $R m $ large and $ Lm$
  finite, the operator $e^{\FF_\tor}$ can be expanded in a series in
  $e^{-\teb} \sim e^{- mR}$ with leading term $e^{\FF_\cyl}$,
  \bee e^{ \mathbf{F} _\tor} &=e^{\mathbf{F}_\cyl } +
  e^{\mathbf{F}_\cyl}\oint\limits _{ \mathcal{C} _{\mathbb{R} }}
  {d\theta \over 2\pi i} \( e^{-\teb (\theta )} - {1\over 2} e^{ -
  2\teb (\theta )}\) \partial_\theta \log\left( 1 + e^{ i\tphib (
  \theta )} \right) \\
  & + {1\over 2!}e^{\mathbf{F}_\cyl} \oint\limits _{ \mathcal{C}
  _{\mathbb{R} }\times \mathcal{C} _{\mathbb{R} }} {d\theta \over 2\pi
  i} {d\theta '\over 2\pi i} e^{ -\teb (\theta )}\, e^{ -\teb (\theta
  ')} \partial_\theta \log\left( 1 + e^{i\tphib ( \theta )} \right)
  \partial_{\theta '} \log\left( 1 + e^{i\tphib ( \theta ')} \right)
  +...  .
   \eee
The integral in the subleading order ($\sim e^{-\teb} $), evaluated by
residues, gives the sum of one-particle excited states in the cross
channel whose rapidity $ \theta $ are determined by the poles of the
integrand,
\be \left\langle e^{\mathbf{F}_\cyl}\, e^{ -\teb ( \theta )} \left( 1
+ e^{ i\tphib ( \theta )} \right) \right\rangle =0 \quad \Rightarrow
\th .  \ee
 The next order ($\sim e^{- 2\teb}$) consists of two terms
 representing a single and a double integral.  The double integral
 gives a sum over the two-particle excited states characterised by
 pairs of residues at $\theta _1$ and $\theta _2$ such that
\be \la{spectrumExc} \left\langle e^{\mathbf{F}_\cyl }e^{ -\teb
(\theta _1) -\teb (\theta _2)} \left( 1 + e^{ i\tphib ( \theta _1)}
\right) \right\rangle =\left\langle e^{\mathbf{F}_\cyl }e^{ -\teb
(\theta_1 ) -\teb (\theta _2)} \left( 1 + e^{ i\tphib ( \theta _2)}
\right) \right\rangle =0.  \ee
The contribution of the unphysical excited state with $\th_1=\th_2$ is
compensated by the single integral.
 
 In the sector with $N$-particle excited states, the spectrum of the
 rapidities $\th_1,...,\th_N$ is determined by the positions of the
 $N$ poles of the integrand, which are determined by the conditions
\bee 
\la{exactTBA}
\tilde \phi _j\equiv { \< \tphib(\th_j) \prod_{k=1}^N e^{-
\teb(\th_k)}\ e^{\FF_\tor}\> \over \< \prod_{k=1}^N e^{- \teb(\th_k)}\
e^{\FF_\tor}\>} = 2\pi I_k +\pi,\quad k=1,..., N. \eee
 with $I_k$   integer numbers.  The explicit expressions for the
 phases $\tilde \phi _k$ are
\bee \la{qqcond} \tilde \phi _j= mL p(\th_j )- i \sum_{k=1}^N \log
S(\th_j-\th_k) + \int {d\th\over 2\pi} \log(1+ e^{-\e(\th )})\
\p_{\th} \log W(\th +\th_j) \eee
where the pseudoenergy $\e(\th)$ satisfies the integral equation
 \bee \la{Widepsl} \e(\theta ) &={ \< \eb(\th) \prod_{k=1}^N e^{-
 \teb(\th_k)}\ e^{\FF_\tor}\> \over \< \prod_{k=1}^N e^{-
 \teb(\th_k)}\ e^{\FF_\tor}\>} \\
&= R E(\theta ) + \sum _{k=1}^N\log W(\theta +\theta _k) - \int
_\mathbb{R} {d\theta '\over 2\pi} K(\theta -\theta ')\log (1+ e^{-
\e(\theta ')}). 
 \eee
The partition function in presence of the $N$-particle excited state
with quantum numbers $I_1,..., I_N$ is
\bee \CZ _{I_1,..., I_N} &= \prod_{j}e^{- R E(\th_j)} \ e^{ R \int {d
p(\theta ) \over 2\pi } \log (1+e^{-\e(\theta )}) } 
\qquad (I_1\le...\le I_N).  \eee
Although the quantisation numbers can coincide, the rapidities of the
multi-particle excited states are all different, $\th_1<...<\th_N$,
because of the cancellations discussed above.  The quantisation
conditions \re{qqcond}-\re{Widepsl} were derived for the sinh-Gordon
model by analytical continuations of the ground state TBA in
\cite{Bajnok:2019yik}, and using a lattice realisation of the theory
in \cite{Teschner:2008ab}.

\subsection{Oscilator representation  of the cylinder and 
 torus partition functions}
\la{sec:complexscf}

 In this section, the HS fields will be given a standard
 representation in terms of creation and annihilation operators of
 bose and fermi type acting in a Fock space.  The bosonic fields $\{
 \eb , \phib \} $ and their mirrors $ \{ \teb , \tphib \}$ will be
 expressed in terms of a free complex chiral boson $\{\bvp,\bbvp\}$ as
 \bee \la{epphivp} \eb(\th) &= \bvp(\th),\quad \ \ \ \phib(\th) = i
 \bbvp(\th -i\pi/2), \\
\teb(\th) & = \bbvp(-\th), \quad \tphib(\th) = i \bvp(-\th + i \pi/2).
\eee
  The fermionic fields $\{\etab, \bpsi\} $ and their mirrors
  $\{\tetab,\tbpsi\}$ will be expressed in terms of a free complex
  chiral fermion $\{\bpsi, \bbpsi\}$ as
 \bee \la{epphivpfer} \bxi(\th) &= \bpsi(\th),\quad \ \ \ \etab(\th)
 =i\bbpsi(\th -i\pi/2), \\
\tbxi(\th) & = \bbpsi( -\th), \quad \tetab(\th) = i \bpsi(-\th + i
\pi/2).  \eee

\medskip

\noindent $\bullet$ {\it Fock space}

\medskip

	\noindent The bosonic field is defined by the mode expansion at
	$\th\to \pm \infty$
\bee \la{chirbos1a} \bvp (\th) &= \sum_{n \ \mathrm{odd}} \JJ _{ n}\
{e^{-n\th} \over n} \,, \quad \bbvp (\th) = \sum_{n \ \mathrm{odd}}
\bar\JJ _{ n} \, { e^{-n\th}\over n} \, , \eee
with the operator amplitudes satisfying the commutation relations
 \be \la{ccrbos} [\JJ_n,\JJT_m] = -n W_n\d_{m+n, 0}\, ,\qquad
 [\JJ_m,\JJ_n] = [\JJT_m,\JJT_n] =0 \qquad (n,m = \text{odd}) \ee
    in which the coefficients $ \W_n $ are determined by the expansion of
 the function $\log \W(\th)$ at $\th \to \pm \infty $,
 \bee \la{expanW} \log W(\th)&= \sum_{k\ge 1, \text{odd}}^\infty
 {W_{n} \over n} \, e^{- n \th} \qquad (\Re\th >0) \\
 &=   \sum_{k\ge 1, \text{odd}}^\infty {W_{n} \over n} \, e^{  n \th}
 \qquad (\Re\th <0)
 .
 \eee
 This is the general form of the expansion for a purely elastic
 scattering matrices \cite{Klassen:1989ui}.  The left and right Fock
 vacua are defined by
   \be \la{defFockvac} \<0|0\> =1, \quad \< 0| \JJ_{-n}=\< 0| \bar
   \JJ_{-n}= \JJ_n|0\rangle = \bar \JJ_n|0\rangle =0 \quad \ \ (n>0) .
   \ee

 Similarly, the fermionic field  is defined by 
  \bee \la{chirbos1af} \bpsi (\th) &= \sum_{n \ \mathrm{odd}} \bb _{
  n}\ {e^{-n\th} \over n} \,, \quad \bbpsi (\th) = \sum_{n \
  \mathrm{odd}} \bbb_{ n} \, { e^{-n\th}\over n} \, , \eee
 %
   \be \la{ccrbosaf} [\bb_n,\bbb_m]_+= -n W_n\d_{m+n, 0}\, ,\qquad
   [\bb_m,\bb_n] _+= [ \bbb_m, \bbb_n] _+=0 \qquad (n,m =
   \text{odd}) \ee
	 \be \la{defFockvacaf} \<0|0\> =1, \quad \< 0| \bb_{-n}=\< 0|
	 \bbb_{-n}= \bb_n|0\rangle = \bbb_n|0\rangle =0 \quad \ \ (n>0) .
	 \ee

\medskip

\noindent $\bullet$ {\it Two-point functions}

\medskip

 \noindent The commutation relations \re{ccrbos} and \re{ccrbosaf} are
 designed so that
 \be \la{twoptfc} \<0| \bvp(\th)\bbvp(\th')|0\> = \<0|
 \bpsi(\th)\bbpsi(\th')|0\> = -\sum_{n\ge 1, \text{odd}} {W_n\over n}
 e^{ - n(\th+\th')}= - \log W(\th+\th').  \ee
The two-point function for the bosons does not depend on the order of
the operators,
 \be \la{symcorp} \<0| \bvp(\th)\bbvp(\th')|0\> =\<0|
 \bbvp(\th')\bvp(\th)|0\> .  \ee
 
 \medskip
 
\medskip

\noindent $\bullet$ {\it Asymptotics at infinity }

\medskip

 \noindent In order to impose the expectation values \re{expvalaux},
 the two Fock vacua are rotated by the evolution operators with
 `times' proportional to the two periods,
  \bee\la{HpHm} & |0\> \to e^{-\HH_-} |0\>\, , \qquad \HH_- = t_{-1}
  \JJ _{ -1} +\bar t_{-1} \JJT _{ -1} \\
  &\< 0|\to \<0| e^{ \HH_+} , \qquad \HH_+ = t_1 \JJ _{ 1} + \bar
  t_{1}\JJT_{ 1} , \\
  &t_{1} = t_{-1} = {Rm \over 2\W_1}\,, \qquad \bar t_{1}=\bar t_{-1}=
  {Lm \over 2 \W_1 }\, .  \eee
More general Hamiltonians,
 \bee \HH_- &= \sum_{j =1}^\infty \(t_{-2j +1} \JJ _{ -2j +1} +\bar
 t_{-2j +1} \JJT _{ -2j +1} \), \\
 \HH_+  &=  \sum_{j =1}^\infty\( t_{2j -1} \JJ  _{ 2j -1} +
  \bar t_{2j -1}\JJT_{2j - 1} \),
\eee
introduce chemical potentials coupled to higher conserved quantities
both in the direct and in the cross channels.  Expectation values
\be \< 0| \bvp(\th)|0\> = \sum_{j\in\IZ} \bar g_{2j-1} e^{-(2j-1)\th},
\quad \< 0| \bbvp(\th)|0\> = \sum_{j\in\IZ} g_{2j-1} e^{-(2j-1)\th} .
\ee
or, in terms of the original HS fields,
 \bee \la{expvalauxGGI} \langle \eb (\theta)\rangle &= \sum_{j\in\IZ}
 \bar g_{2j-1} e^{(2j-1)\th}, \ \ \ \langle \phib (\theta)\rangle =
 \sum_{j\in\IZ} (-1)^{ j-1} \ g_{2j-1} e^{(2j-1)\th} ,\\
	  \langle \teb (\theta)\rangle &= \sum_{j\in\IZ} g_{2j-1}
	  e^{-(2j-1)\th},\ \ \ \langle \tphib (\theta)\rangle =
	  \sum_{j\in\IZ} (-1)^{ j } \ \bar g_{2j-1} e^{-(2j-1)\th}\, \eee
with $g_{n} , \bar g_{n}\ge 0$, are generated by choosing the `times'
as
\bee t_{2j -1} &= { g_{2j-1}\over W_{2j -1}} ,\quad \bar t_{2j -1} = {
\bar g_{2j-1}\over W_{2j -1}} .  \eee

\medskip

\medskip

\noindent
$\bullet$ {\it    The cylinder partition function }

\medskip

 \noindent The partition function on a cylinder, eq.
 \re{partfGENbisTBA}, takes the form of a Fock-space expectation
 value,
 \bee \la{partfGENbisTBAosc} \mathcal{Z}_\cyl^ { (L ) }
 \Big|_{\e_0=0}&= { {\langle0| e ^{ \HH _+} \ e^{ \mathbf{F} _\cyl } \
 e ^{ -\HH _-} |0\rangle} } , \\
  \\
	\FF_\cyl &= \int\limits_{\mathbb{R} } {d\th \over 2\pi i} \left[
	-\s \log\left( 1 -\s\, e^{- \bvp }\right) \p_\th \bbvp^\mm + {
	\bpsi \over 1-\s e^{ {\bvp } } } \p_\th \, \bbpsi^\mm \right] .
\eee
Here the following notation was used for functions with shifted
arguments,
\be f^{[\pm]}(\th)\equiv f(\th \pm i\pi/2 \mp i 0).  \ee
The expectation value \re{partfGENbisTBAosc} corresponds to certain
normalisation of the infinite-volume free energy $\CF_0=-LR \e_0$,
namely

  \be \la{defepnul} \CF_0 = - \sum_{j\in\IZ} {|2j-1 |\over W_{2j-1}}
  g_{2j-1} \bar g_{-2j+1} =-RL {m^2 \over 2 W_1} + ...  . \ee
   This normalisation for $\e_0$, namely
   \be \la{norminfven} \e_0= {m^2 \over 2 W_1} , \ee
is the natural one in the following sense.  The $LR$ term, which
reflects the dynamics in infinite spacetime, appears only in the
large-volume expansion of the full free energy $\CF_\cyl^{(L,R)}$, and
not in the small-volume expansion.\footnote{I thank S. Lukyanov for
having instructed me about that.  } In the normalisation $\e_0=0$, the
bulk vacuum energy appears (with opposite sign) in the small-volume
expansion of the effective central charge (see e.g. section 20 of
\cite{Mussardo-Book}).
 
The ground-state energy in the cross channel is minus the logarithmic
derivatives of the partition function with respect to $R$,
  \bee \la{derivativeinR} \mathcal{E}(L)\equiv - { \p\log\CZ_\cyl
  ^{(L,R)}\over \p R}&= {m\over 2 W_1} { {\langle0| e ^{ \HH _+} \ \(
  \JJ_1 e^{\FF_\cyl} - e^{\FF_\cyl} \JJ_{-1} \)e ^{ -\HH _-}
  |0\rangle} \over {\langle0| e ^{ \HH _+} \ e^{ \mathbf{F} _\cyl} \ e
  ^{ -\HH _-} |0\rangle} } \\
 &= L \e_0 +\s\int {dp(\th)\over 2\pi } \ \ \log\left( 1 -\s\, e^{- \e
 }\right), \quad \ \eee
with $\e=\<\bvp\>_\cyl$ determined by \re{Widentity}.

\medskip

\medskip
\medskip

\noindent
$\bullet$ {\it    The torus partition function }

\medskip

\noindent
  With the above  normalisation for $\e_0$, the
  partition function on the
  torus  is given by the vacuum expectation value
 \bee \la{partfGENbis} \mathcal{Z}_\tor^ { (L,R ) }&= {\langle0| e ^{
  \HH _+} \ e^{ \mathbf{F} _\tor } \ e ^{ -\HH _-} |0\rangle} 
   \eee
with $\FF_\tor$ given either by
\bee \la{OperFFosc11} \FF_\tor &= \s \int\limits_{\mathbb{R} } {d\th
\over 2\pi i } \left[ \log\left( 1 -\s\, e^{- \bvp }\right) \, \p_\th
\bbvp ^\mm+ { \bbpsi ^ \mm \, \p_\th \bpsi \over 1-\s e^{ {\bvp } } }
\, \right] \\
	 & -\s \oint\limits_{ \CC_{\IR} } {d\th \over 2\pi i } {1\over 1-
	 \s e^{ \bvp ^\pp }} \left[ { \log\left( 1- \s\, e^{ - \bbvp }
	 \right) }\, \p_\th \bvp ^\pp + { \bpsi ^\pp\, \partial _\th
	 \bbpsi \over 1- \s\, e^{ - \bbvp } } \right] , \eee
which corresponds to   \re{GFTcfirstnew1}, or by the
mirror-transformed expression
   \bee \la{OperFFosc22} \FF_\tor &= \s \int\limits_{\mathbb{R} }
   {d\th \over 2\pi i } \left[ \log\left( 1 -\s\, e^{- \bbvp }\right)
   \, \p_\th \bvp ^ \pp+ { \bpsi ^\pp \, \p_\th \bbpsi \over 1-\s e^{
   {\bvp } } } \, \right] \\
& -\s \oint _{ \CC_{\IR} } {d\th \over 2\pi i } {1\over 1- \s e^{
\bbvp ^ \mm } } \left[ \log\left( 1- \s\, e^{ - \bvp } \right) \,
\p_\th \bbvp ^\mm + { \bbpsi ^\mm\, \partial _\th \bpsi \over 1-\s e^{
-{\bvp } } } \right] , \eee
 which  corresponds  to   \re{GFTcfirstnew2}.

With the choice \re{OperFFosc11}, one obtains for the energy at finite
volume $L$ and finite temperature $ 1/R$
  \bee \la{derivativeinRtor} \mathcal{E}(L,R)= { \p\log\CZ_\tor
  ^{(L,R)}\over \p R}&= {m\over 2 W_1} { {\langle0| e ^{ \HH _+} \ \(
  \JJ_1 e^{\FF_\tor} - e^{\FF_\tor} \JJ_{-1} \)e ^{ -\HH _-}
  |0\rangle} \over {\langle0| e ^{ \HH _+} \ e^{ \mathbf{F} _\tor } \
  e ^{ -\HH _-} |0\rangle} } \\
&= L \e_0 +\s\int {dp(\th)\over 2\pi } \ \ \<\log\left( 1 -\s\, e^{-
\bvp }\right) \>_\tor \\
  & -\s \oint\limits_{ \CC_{\IR} } {d\th \over 2\pi } \ p(\th) \left\<
  { \p_\th \bvp ^\pp + { \bpsi ^\pp\, \partial _\th \bbpsi \over 1-
  \s\, e^{ - \bbvp } }\over (1- \s e^{ \bvp ^\pp }) (1- \s\, e^{ \bbvp
  } ) } \right\>_\tor, \eee
where $\<\CO\>_\tor$ is the normalised expectation value in the
ensemble of loops \footnote{By the property \re{symcorp},
for the operators in \re{derivativeinRtor}   it is not
important whether the operator is inserted before or after the
exponential.}
 \be \<\CO\>_\tor\equiv { \langle0| e ^{ \HH _+} \ \CO\ e^{ \mathbf{F}
 _\tor } \ e ^{ -\HH _-} |0\rangle \over \langle0| e ^{ \HH _+} \ e^{
 \mathbf{F} _\tor } \ e ^{ -\HH _-} |0\rangle}.  \ee

\subsection{Example: the sinh-Gordon model }
\la{section:sinh-G}
 \def\YY{{Y\!\!\!  \!Y}}

The Sinh-Gordon model, described by the Euclidean action
 \be \la{actionshG} \mathcal{A} = \int d^2 x \left[ {1\over 4\pi}
 (\nabla\phi)^2 + {2\mu } \cosh ( 2b\phi)\right], \ee
is the simplest interacting scattering theory
\cite{Vergeles:1976ra,Arefy-Kor}.  Its spectrum consists of one
neutral particle whose mass $m$ is determined by the coupling constant
$\mu$ and the dimensionless parameter $b$ \cite{Zamolodchikov:1995xk}.
The scattering factor $S(\th-\th')$ for two physical (or two mirror)
particles corresponds to fermionic TBA statistics, $\s= S(0)=-1$, and
is given by \cite{ARINSHTEIN1979389}
  \bee \la{smatsG} 
  S(\th) &= \frac{\sinh (\theta )-i \sin (\pi \alpha
  )}{\sinh (\theta )+i \sin (\pi \alpha )} 
  \\
   & = 
   \frac{\tanh \left(\frac{1}{2} (\theta -i \pi  \alpha )\right)}{\tanh \left(\frac{1}{2} (\theta +i \pi  \alpha )\right)}\eee
 where the parameter $\a$ and the coupling $b$ are related by
  \be \a= {b^2\over 1+b^2} .  \ee
For one physical and one mirror particle, the scattering factor is
$W_\a(\th+\th')$, with
\be W(\th)= \frac{\cosh (\theta )+\sin (\pi \alpha ) }{\cosh (\theta
)-\sin (\pi \alpha )}.  \ee

The mass and the $S$-matrix are invariant under the strong/weak
coupling duality transformation $b\to 1/b$, or $\a\to 1-\a$.  For the
parameter
	 \be a = 1- 2 \a ={1-b^2\over 1+b^2 } \ee
 used in \cite{Zamolodchikov:2000kt}, the duality acts as $a\to - a$.

The oscillator representation of the torus partition function is given
by the general expression \re{partfGENbis}, with the commutation
relations for the bosonic and fermionic oscillators, eqs.  \re{ccrbos}
and \re{ccrbosaf}, determined by the series expansion
 \bee \log W(\th) &=\sum _{n\ge 1, \text{odd}} { W_n\over n} \
 e^{-n\th},\quad W_n= 4\cos{n\pi a\over 2}.  \eee
The large volume asymptotics, eq.  \re{defepnul}, matches precisely
the perturbative result derived in \cite{DESTRI1991251} by assuming
normal ordering in the interaction potential in \re{actionshG},
 \be \e_0 = {m^2 \over 2 W_1} = {m^2\over 8 \sin \pi \a}.  \ee
This normalisation of $\e_0$ can be also obtained from the requirement
that there is no $m^2 LR$ term in the small-volume expansion.  It is
perhaps worth trying to find the massless limit of the torus partition
function of sinh-Gordon.

The Ward identity \re{Widentitya} can be given a functional form using
the properties of the scattering kernel for the S-matrix \re{smatsG}.
The scattering kernel decomposes as decomposes as
  \bee \la{KKcirc} K( \th)& ={1\over i} \( {1\over \sinh (\th -
  i\pi\a)} - {1\over \sinh (\th + i\pi\a)}\) = {1\over \cosh(\th +i\pi
  a/2) }+{1\over \cosh(\th -i\pi a/2) }.  \eee
 Furthermore, the `universal' kernel $1/\cosh\th$ satisfy the identity
 \be\la{diffeq} \la{KoDiLa1} {1\over \cosh(\th+i\pi /2)} + {1\over
 \cosh(\th-i\pi /2)} = 2 \d(\th).  \ee
These properties have been used in
\cite{Zamolodchikov:2000kt,Lukyanov:2000jp} to derive functional
equations for the pseudoenergy in the TBA limit.

With the shift operator on the lhs of \re{KoDiLa1} applied to
\re{Widentitya} , the integral in the first term becomes the shift
operator in \re{KKcirc} applied to the integrand, while the second
term simply vanishes because the integration contour is not on the
real axis.  As a result, the Ward identity takes a functional form,
  \bee \la{WWI1} \< \eb^\pp (\th) +\eb^\mm (\th) \>_\tor &=
  \<\log(1+e^{-\eb ^{[a]} (\th) }) + \log(1+e^{-\eb ^{[-a]} (\th) })
  \>_\tor.  \eee
 where  the   shorthand notation for 
  the shift  by $i\g\pi/2$,
 \be \la{defshiftsf} f^{[\g]}(\th)\equiv f(\th + \hf i \pi \g) \qquad
 -\pi<\g< \pi.  \ee
 A similar functional equation can be derived for $\teb$ starting with
 the dual representation \re{GFTcfirstnew2} of the operator
 $\FF_\tor$,
   \bee \la{WWI2} \< \teb^\pp (\th) +\teb^\mm (\th) \>_\tor &=
   \<\log(1+e^{-\teb ^{[a]} (\th) }) + \log(1+e^{-\teb ^{[-a]} (\th)
   }) \>_\tor .  \eee

In the TBA limits $Rm\gg 1$ or $Lm\gg1$, these Ward identities become
functional equations for the Y-function $ Y(\th) = e^{-\e(\th)} $,
 \be \la{Ysys} \la{Ysisshort} Y^{[1]}Y^{[-1]} =(1+ Y^{[a]})(1+
 Y^{[-a]}), \ee
 which hold in a small strip around the real axis.  It was shown in
 \cite{Zamolodchikov:2000kt} that this relation can be continued to
 the whole rapidity plane and that $Y(\th)$ is entire function of
 $\th$ with essential singularities at $\th\to\pm\infty$.
 Furthermore, the Q-function defined by
 \be Y = Q ^{[a]} Q^{[-a]} \ee
  satisfies the quadratic functional identity
  \cite{Zamolodchikov:2000kt,Lukyanov:2000jp}
 \bee \la{Xsys} Q ^{[1]} Q ^{[-1]}- Q ^{[a]} Q^{[-a]} =1.  \eee
The lhs of this identity can be interpreted \cite{Negro:2013wga} as
the quantum Wronskian of the two independent solutions $Q(\th)$ and $
\hat Q(\th)= Q^{[a-1]}=Q(\th + i \pi {a-1\over 2}) $ of Baxter's T-Q
relation,
 \be \la{TQsys} T Q = Q^{[1+a]} +Q^{[-1-a]},\ \ T \hat Q = \hat
 Q^{[1+a]}+ \hat Q^{[-1-a]} , \qquad \left| \begin{matrix} \hat Q^{[a]}
 & Q^{[a]} \\
      \hat Q^{[-1]}&  Q^{[-1]} \\
   \end{matrix}\right|=1
.
\ee
%
By a standard argument, the function $T(\th)$ enjoys the periodicity
property $T^{[1-a]}=T$ and is entire function of the variable $\zeta=
e^{4\th/(1-a)}$ defined by convergent series in $\zeta $
\cite{Zamolodchikov:2000kt}.  A dual object $\tilde T(\th) $ with
periodicity $\tilde T^{[1+a]}=\tilde T$ is constructed in the same way, with $a$
replaced by $ -a$.

The functional equations \re{Ysys}, as well as \re{Ysys} and
\re{Xsys}, hold also in presence of an excited eigenstate of the
finite-volume Hamiltonian.  By the cross analyticity $S(\th)S(\th +
i\pi)=1$, the integral equation \re{Widepsl} leads to the same
Y-system \re{Ysys} as for the ground state.  In this sense, the
complete spectrum of excitations and therefore the partition function
on the torus can be in principle inferred from the functional
equations.  However this does not mean that the Ward identities
\re{WWI1} and \re{WWI2} can be transformed into functional equations.
The functional equations hold only for eigenstates of the
finite-volume Hamiltonian, and not for a linear combination of them.
\section{ Summary and discussion}
\la{section:conclusion}

The main message I want to convey in this paper is
 that the partition function of a
massive integrable QFT on the torus can be mapped to a gas of loops
with two-body interaction determined by the scattering data.  
The loop-gas  can be formulated for any theory
with factorised scattering.    
This paper was focused on the case of diagonal scattering 
and no bound states, 
where the partition function was formulated as an
effective field theory.  The effective fields are the HS  fields
for the Hubbard-Stratonovich transformation introduced to decouple the
two-body interaction of loops.

In the limit where one of the periods of the torus becomes large, the
EFT becomes a mean-field type and gives a field-theoretical
formulation of the Thermodynamical Bethe Ansatz.  When both periods
are finite, there is no reason to expect that the effective QFT can be
solved exactly; it can be rather used to compute the exponential
corrections to the mean-field limit.  The leading and the sub-leading
exponential corrections in the long period $R$ can be obtained by
iterating the mean-field equations \re{Widentityator}.

The EFT can be also formulated for the correlation functions of local
operators on a cylinder or on a torus.  Depending on the observable, a HS field is associated with
each homology cycle.  The attractiveness of this approach is that it
does not require a regularisation procedure to extract the
finite-volume observables given the infinite-volume ones.  In case of
the one-point functions on a cylinder, the EFT gives immediately
Leclair-Mussardo formula, eq.  \re{LMseries}.  In the case of the
torus, the LM series will include world lines in both physical and
mirror kinematics.  In general, any correlation function of local
operators can be formulated in terms of the ensemble of loops. 

Another worldsheet geometry for which the EFT can be developed is that
of a finite cylinder.  The derivation of the EFT in this case
essentially repeats that for the torus.  The first winding number will
be the number of times the loop winds around the $L$-cycle, while the
second winding number will count the number of reflections from the
two boundaries.  When the distance between the two boundaries is
finite, the two boundaries start to see each other and boundary
entropy does not factorise into a product of two $g$-functions, as it
does in the cylinder limit.

 It is straightforward to generalise the loop-gas
description and the EFT  to the case of
purely elastic scattering theories involving several particles.  There
will be a pair of HS fields associated with each particle species.
In the case of a non-diagonal scattering 
and/or bound states of the spectrum, a procedure analogous to the nested Bethe Ansatz  should  be developed.

\bigskip

\no
{\bf Acknowledgements}

\smallskip

\noindent I thank Benjamin Basso for a critical reading of the
manuscript, and Jo\~ao Caetano, Fabian Essler,  Shota
Komatsu, Sergey Lukyanov, Hubert Saleur, Fedor Smirnov and especially
Giuseppe Mussardo for stimulating and enlightening discussions.  The
kind hospitality and support from the Scuola Internazionale per gli
Studi Avanzati (SISSA), Trieste, where part of this work was done, is
highly acknowledged.  This research was supported in part by the
National Science Foundation under Grant No.  NSF PHY-1748958.

\appendix

    \section{ Feynman  graph expansion } 
    \la{sec:Feynman}
     
\subsection{Feynman  rules for the cylinder }

\begin{figure}[h!]
  \hskip 1cm \begin{minipage}[t]{0.89\linewidth}
  \centering    
            \includegraphics[width=11.5 cm]{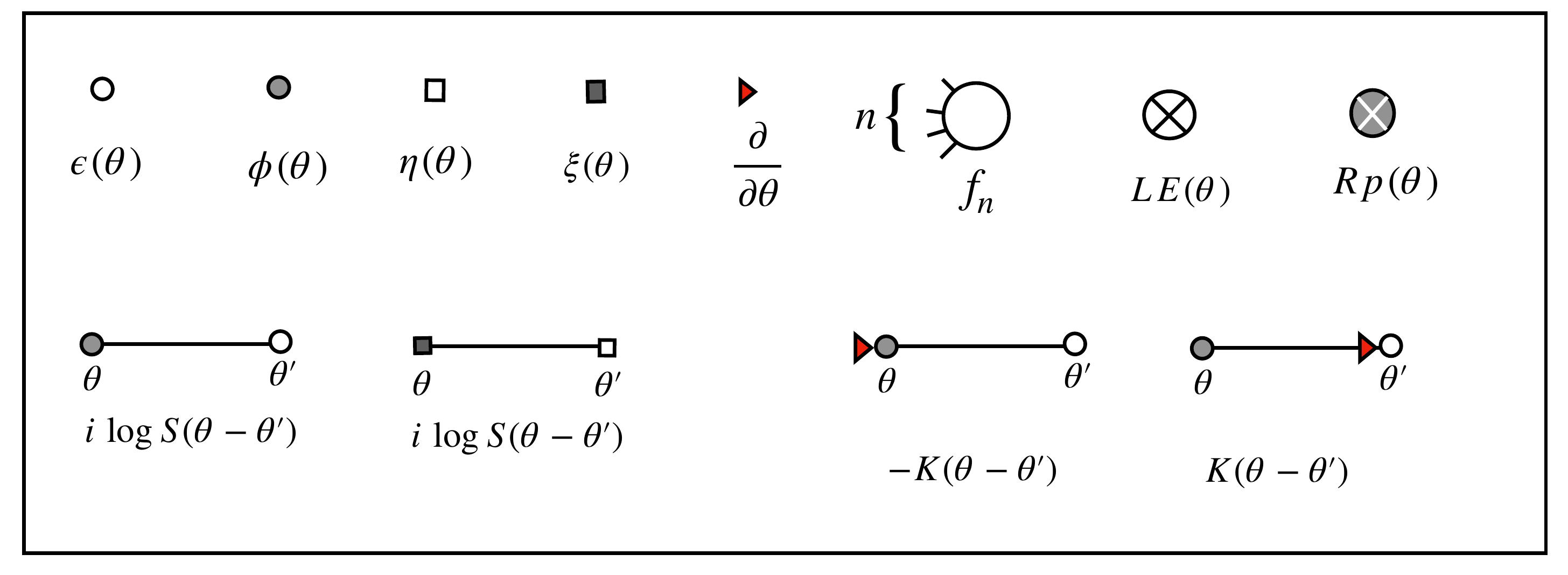}
\vskip - 0.0cm \caption{ \small Feynman rules for the HS fields on a cylinder.  The
coefficients $f_n$ are depicted by white blobs with $n$ legs.  The
fields $\eb, \phib , \etab, \bpsi$ are represented respectively by
light and dark dots and squares.  The derivative $\p_\th$ is
symbolised by a red triangle pointing to the element at which the
derivative is applied.  }
  \label{fig:Feynman-cyl}
  \end{minipage}
  \vskip 0.5 cm
         \centering
	 \begin{minipage}[t]{0.65\linewidth}
            \centering
            \includegraphics[width=9.2 cm]{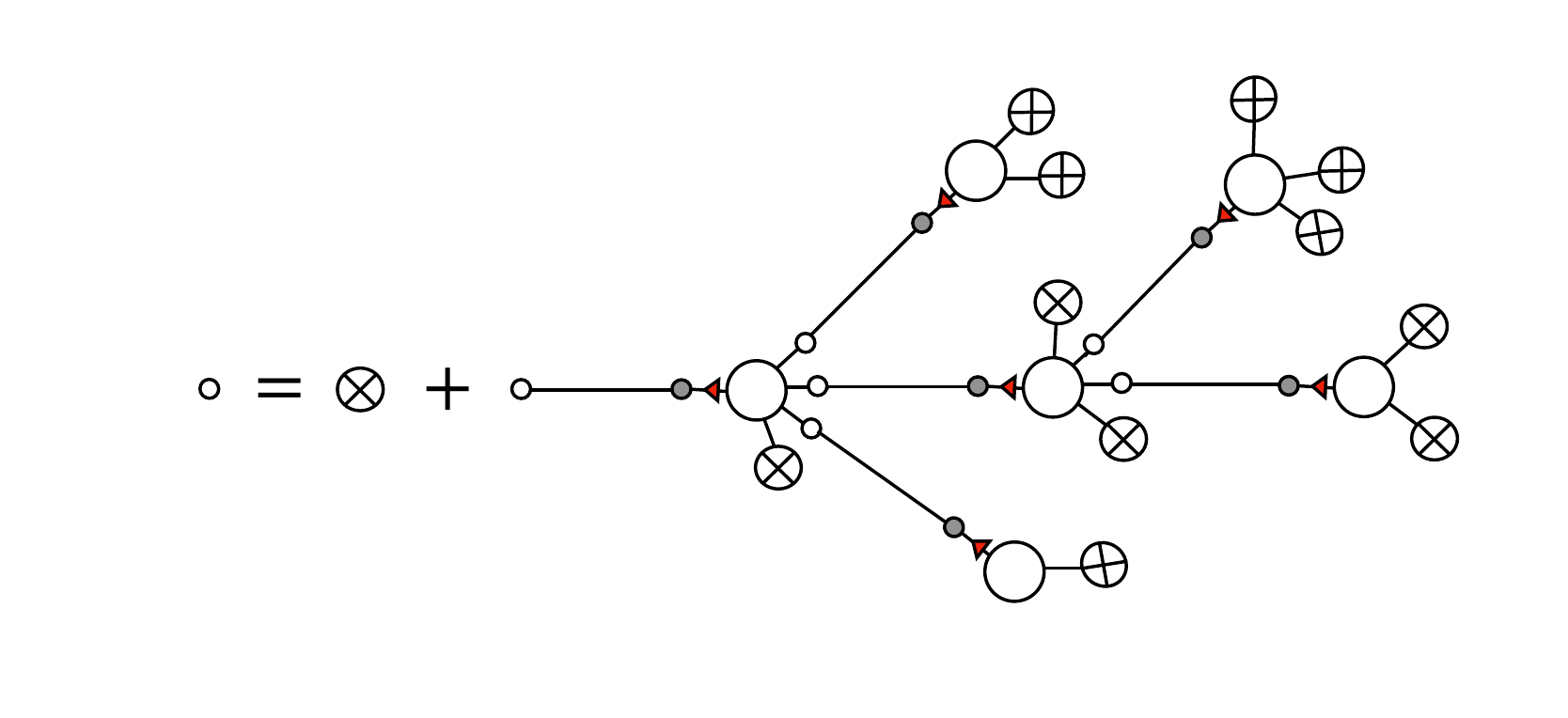}
			   \vskip -0.6cm \caption{ \small Feynman graph expansion
			   for the pseudoenergy $\e= \< \eb\>_\cyl$ }
			   \la{fig:treescyl} \end{minipage}
             	 \begin{minipage}[t]{0.67\linewidth}
	  \vskip 0.5cm
	   \centering
			\includegraphics[height=3.2cm]{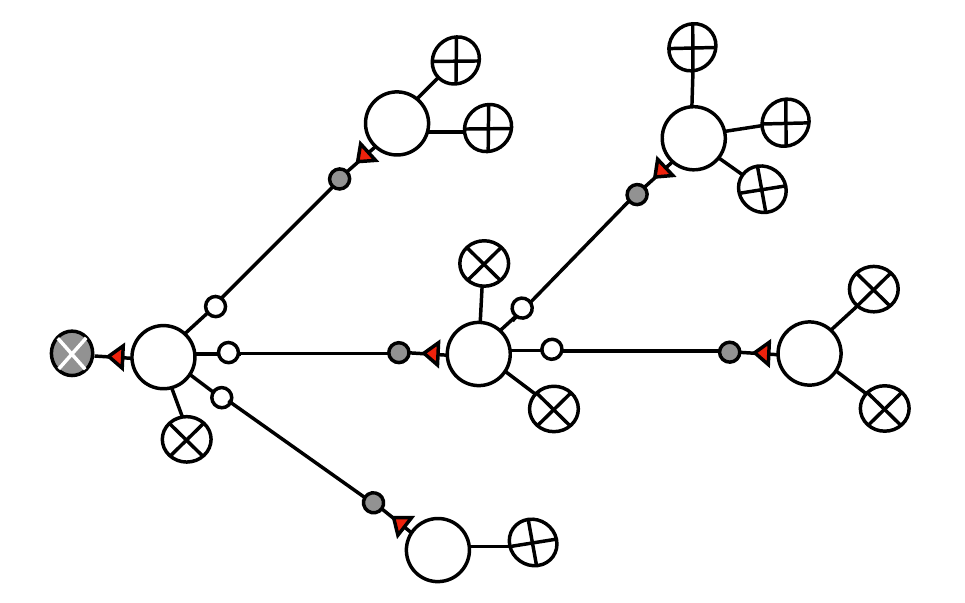}
  \vskip -0.2cm \caption{ \small A typical Feynman graph contributing
  to the free energy }
  \label{fig:ftree}
         \end{minipage}
\centering
\vskip 0.5cm 
              \begin{minipage}[t]{0.59\linewidth}
            \centering
			\includegraphics[width=9.5cm]{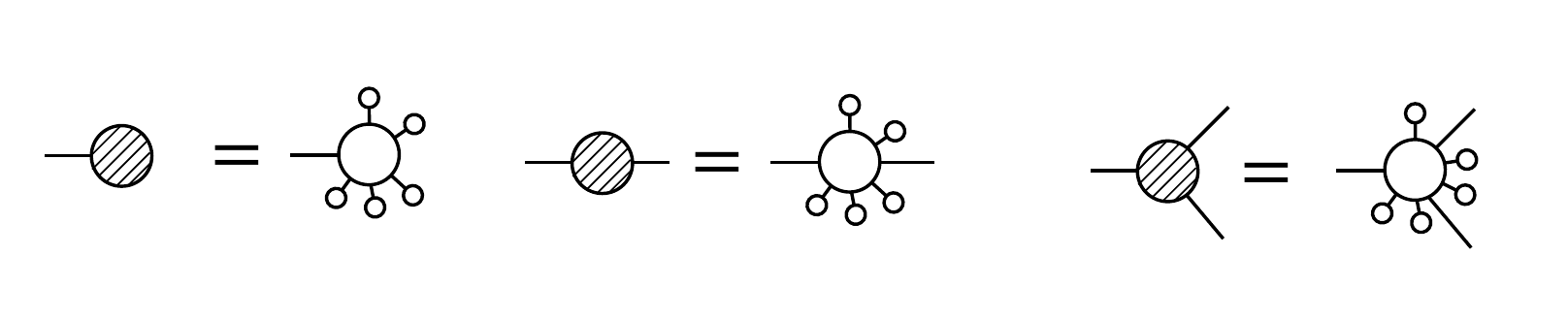}
	  \vskip -0.4cm \caption{ \small The dressed vertices $f_1[\e],
	  f_2[\e], f_3[\e]$.}
\label{fig:dressed}
         \end{minipage}
           \end{figure}

 Feynman diagrams provide a useful scheme to analyse the effective
 QFT. They already appeared in disguise as the tree expansion
 developed in \cite{Kostov:2018ckg,Kostov:2018dmi}.  Let us derive the
 tree expansion assuming fermionic TBA statistics, $\s=-1$, for which
 the interacting potential can be Taylor-expanded around $\eb=0$.  In
 order to derive the diagrammatic rules for the interaction potential
 \re{partfGENbisTBA}, the latter should be expanded in the fields,
\bee \la{Feynmpot} \mathbf{F}_\cyl &= \int_{\mathbb{R} }{d\theta \over
2\pi }\ \left[ \log\left( 1 +e^{- \eb }\right) \partial _\th \phib + {
\etab \partial _\th \bxi \over 1 + e^{ {\eb } } } \right] \\
&= \sum_{n=0}^\infty \int_{\mathbb{R} }{d\theta \over 2\pi }\ \left[
f_{n+1} {\eb^n\over n!} \partial _\th \phib + f_{n+2} {\eb^n\over n!}
\etab \partial _\th \bxi \right] .  \eee
  The coefficients of the expansion,
  \be f_1= \log 2,\ f_2= -\hf, \ f_3=\textstyle{1\over 4}, \ f_4=0, \
  f_5= - \textstyle{1\over 8}, \ f_6=0, \ \dots \ee
  are given by the general formula
\be f_{k+1} & \equiv (-1)^k \left(1-2^k\right) \zeta (1-k)= (-1)^{k-1}
\text{Li}_{1-k}\left(- 1\right).  \ee
The Feynman rules are depicted in fig.  \ref{fig:Feynman-cyl}.  To
avoid using too many pictorial notations, the HS fields and their
(dressed) expectation values will be denoted by the same symbol.  The
precise meaning will be clear from the context.  The expansion for the
interacting potential for the HS fields is represented pictorially as
\vskip -1cm
\be \FF_\cyl = \sum_{n=0}^\infty {\setlength{\unitlength}{3pt}
\begin{picture}(15, 15)(-4,5) \put(-3.5,
+1){\includegraphics[height=1cm]{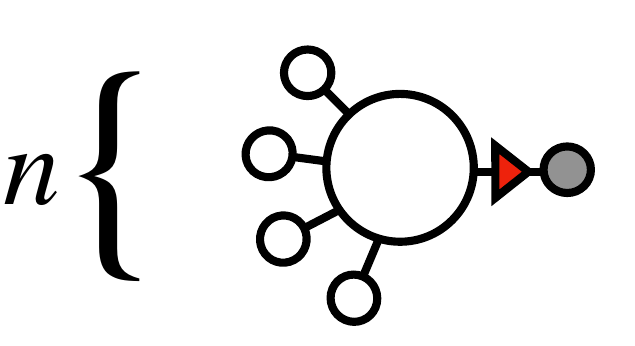}} \end{picture} } \hskip
0.0cm \ \ \ \ + \sum_{n=0}^\infty \ \ \ \ \ \ \ \ \ \ \ \ \ \ \
\setlength{\unitlength}{3pt} \begin{picture}(-15, -15)(9,5) \put(-3.5,
+1){\includegraphics[height=1cm]{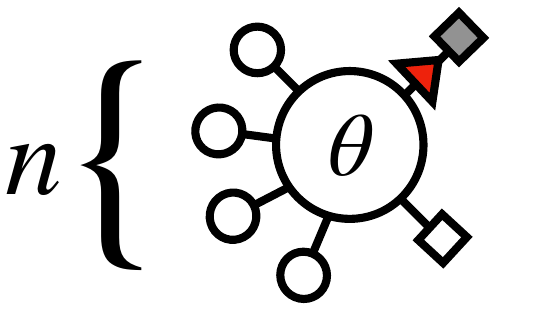}} \end{picture} \hskip 1.7cm
\ \ .  \ee

Since the vertices have only one $\phi$-leg, the vacuum Feynman
diagrams can have at most one cycle.  The diagrams containing
fermionic and bosonic cycles have exactly the same weights but
different signs.  Thus all Feynman graphs are trees.  A typical
Feynman graph contributing to the expectation value $\e=\< \eb\>_\cyl$
is shown in Fig.  \ref{fig:treescyl}.

The Feynman graph expansion can be defined for any classical
background $\e$.  The corresponding Feynmann rules will be the same,
with the vertices and the tadpoles replaced as
\bee f_n\to f_n[\e] \equiv &\sum_{k=0}^\infty f_{k +n} {\e^k\over k!}
= (-1)^{n-1} \text{Li}_{2-n}\left(- e^{-\e}\right), \ \ n= 1,2,...
\eee
This relation between the old, `bare', vertices and the new,
`dressed', ones is illustrated by fig.  \ref{fig:dressed}, where the
dressed vertices are represented by hatched blobs.  The first three
vertices are
   \bee
f_1 [\e]&= \log(1+e^{-\e})
,\ \ 
f_2[\e] 
  = -{1\over 1+ e^\e}
 ,
\quad  f_3[\e]= {1\over (1+e^\e)(1+e^{-\e})}
.
 \eee
If the background $\e$ satisfies the classical equation
\re{Widentity}, the diagram technique does not produce trees.  Since
the loops cancel, the free energy is given by a single graph
\vskip -0.4cm
\bee \CF_\cyl &= R\int {d\th\over 2\pi } f_1[\e(\th)]\p p(\th) =
{\setlength{\unitlength}{3pt} \begin{picture}(5, 0)(-4,5) \put(-3.5,
+3.5){\includegraphics[height=0.5cm]{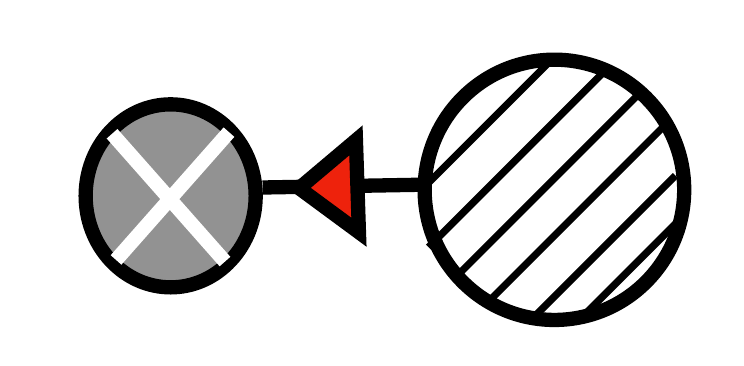}} \end{picture}
} \hskip 0.51cm .  \eee
The  pictorial form of the   Ward identities \re{Widentitya}
and  \re{Widentitya2} is
\bee
 & {\setlength{\unitlength}{3pt} \begin{picture}(5, 0)(-4,5) \put(-3.5,
+3.5){\includegraphics[height=0.5cm]{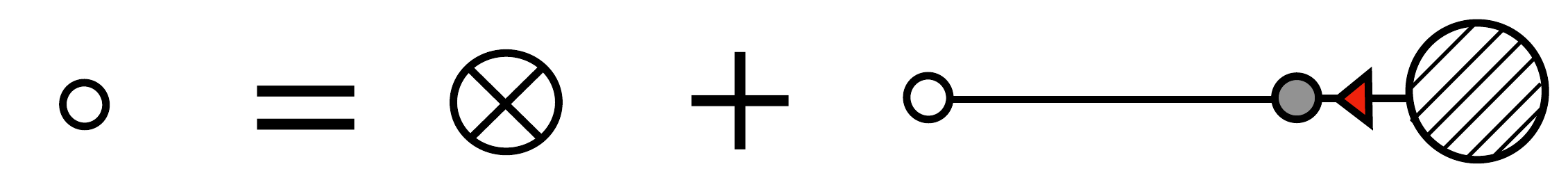}} \end{picture} }
 \hskip 3.61cm  ,\ \ \ 
\\
& {\setlength{\unitlength}{3pt} \begin{picture}(5, 0)(-4,5) \put(-3.5,
+3.5){\includegraphics[height=0.5cm]{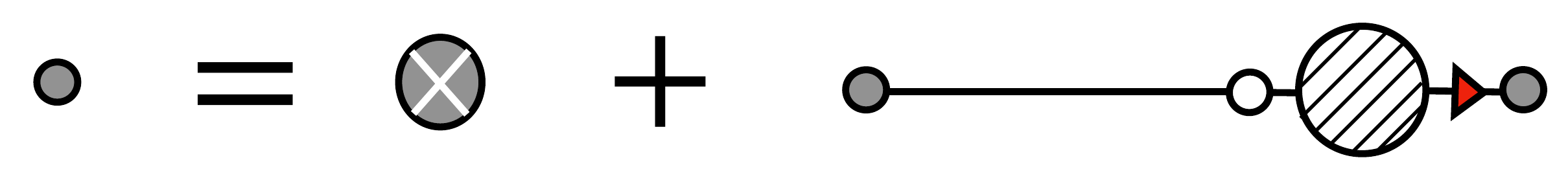}} \end{picture} }
 \hskip 4.51cm 
 \eee
and that of  the Leclair-Mussardo formula  is
 \bigskip \bee \la{LMgr} \ \< \CO \>_R &= \sum_{n=0}^\infty
 {(-1)^n\over n!} \int \prod_{j=1}^n {d\th_j\over 2\pi }
 f_2[\e(\th_j)] \ F^c_{2n}(\th_1, \dots, \th_n) \\
 \\
 &= \sum_{n=0}^\infty (-1)^n {\setlength{\unitlength}{3pt}
 \begin{picture}(5, 0)(-4,5) \put(-3.5,
 -2.5){\includegraphics[height=1.9cm]{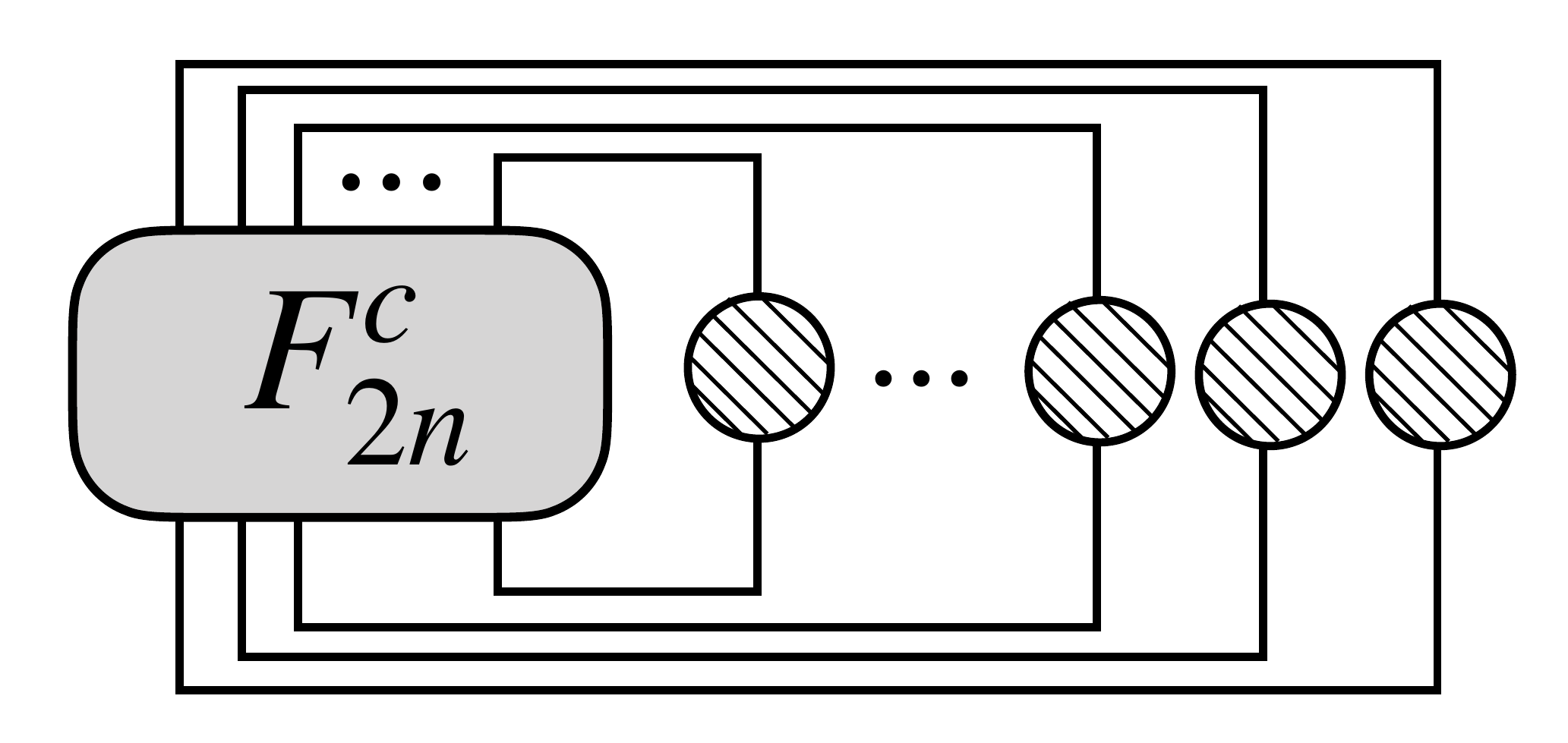}} \end{picture} } \hskip
 1.21cm .  \eee

\subsection{Feynman diagram technique for the torus }

Let us again choose the fermionic TBA statistics, $\s=-1$.  The
interaction potential is expanded as
 \bee \la{GFTcfirstnew1ff} \FF_\tor = & \int_{\mathbb{R} } {d \th
 \over 2\pi } \left[ f_1[ \eb ] \partial _\th \phib \, - f_2[\eb ]
 \etab \p _\th \bxi \right] -\oint _{ \CC_{\IR} } {d \th \over 2\pi }
 f_2[-i\tphib ] \left[ f_1[\teb ] { \partial _\th \tphib } - f_2[\teb
 ] \tetab \partial_\th \tbxi \right] \, \\
&= \sum_{n=0}^\infty \int _\IR {d\th\over 2\pi} \( f_{n+1} \frac{
\eb^{n } }{n!} \p_\th \phib - f_{n+2} \frac{ \eb^{n } }{ n!} \etab \p
_\th \bxi \) \\
 &+ \sum_{n,m=1}^\infty \oint_{ \CC_{\IR} } {d \th \over 2\pi } \(
 -f_{m+1} f_{n+2 } { \teb^{m}\over m!} \ \p_\th\tphib
 {(-i\tphib)^n\over n!} + f_{n+2 } f_{m+2} { \teb^{m }\over m!} {
 (-i\tphib)^n\over n!} \tetab \partial_\th \tbxi \) \eee

   \begin{figure}[h!]
         \centering
	 \begin{minipage}[t]{0.85\linewidth}
            \centering
            \includegraphics[width=12 cm]{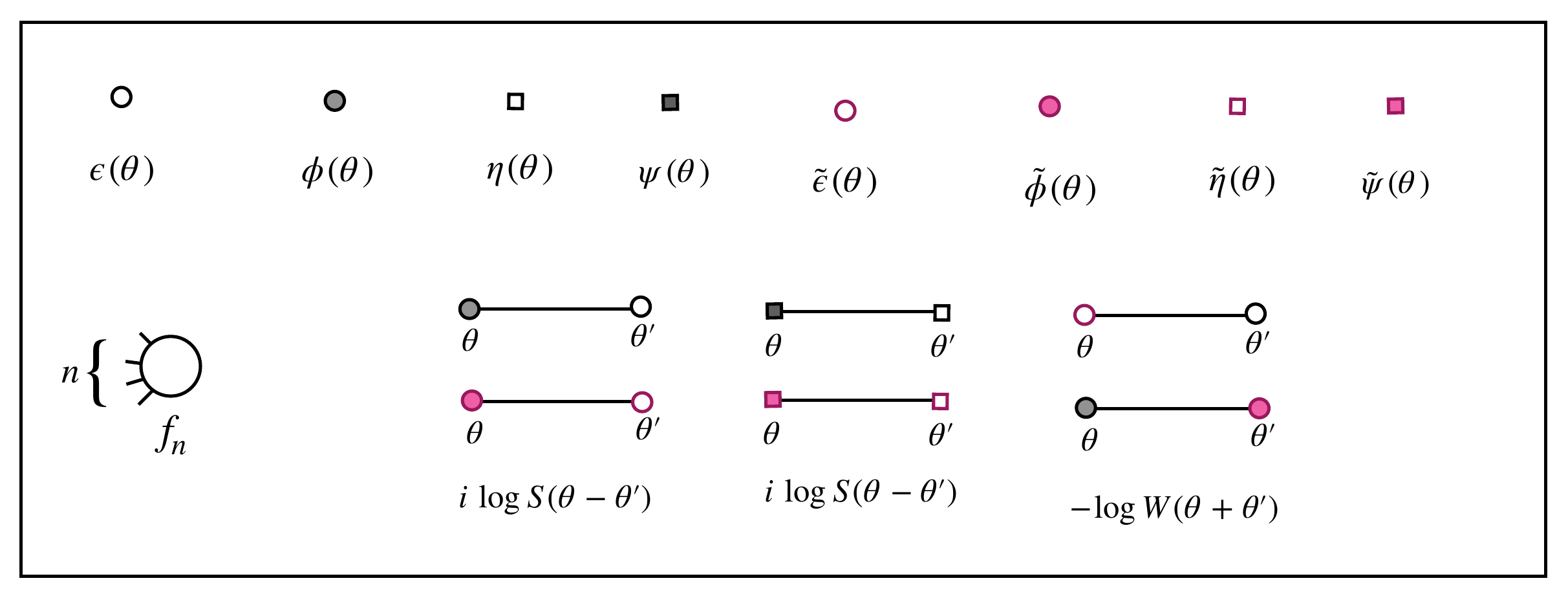}
\vskip - 0.0cm \caption{ \small Feynman rules for the HS fields on  a
torus.  The mirror-transformed fields are represented by the same
symbols colored in magenta.  }
  \label{fig:feynmantor}
  \end{minipage}
 \end{figure}

   \begin{figure}[h!]
           \centering
   \begin{minipage}[t]{0.65\linewidth}
            \centering
            \includegraphics[width=16 cm]{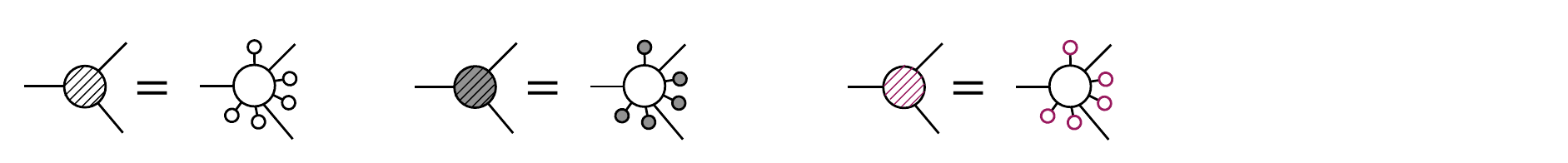}
            \vskip - 0.0cm
\caption{ \small   Dressed vertices for the torus.
   }
  \label{fig:dressedvtor}
  \end{minipage}
 \end{figure}
 \begin{figure}[h!]
           \centering
   \begin{minipage}[t]{0.65\linewidth}
            \centering
			\includegraphics[width=8 cm]{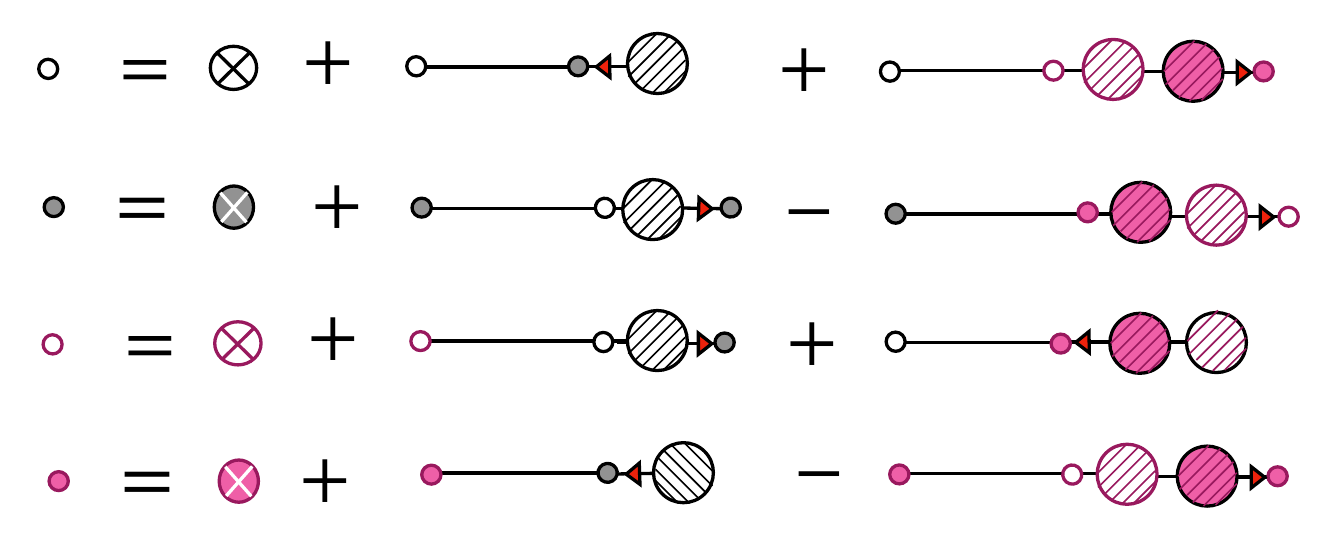}
\vskip - 0.0cm \caption{ \small Pictorial form of the mean-field
equations \re{Widentityator} for the classical values $\e, \phi,
\tilde \e, \tilde \phi$.  }
  \label{fig:meanfieldtor}
  \end{minipage}
 \end{figure}

With the Feynman rules depicted in fig.  \ref{fig:feynmantor}, the
interacting potential for the HS fields can be represented as \vskip
-1cm
  \bee \la{ftorgraphical} \FF_\cyl &= \sum_{n=0}^\infty
  {\setlength{\unitlength}{3pt} \begin{picture}(15, 15)(-4,5)
  \put(-3.5, +2){\includegraphics[height=0.8cm]{int1.pdf}}
  \end{picture} } \hskip 0.0cm \ \ + \sum_{n=0}^\infty \ \ \ \ \ \ \ \
  \ \ \ \ \ \setlength{\unitlength}{3pt} \begin{picture}(-15,
  -15)(9,5) \put(-3.5, +2){\includegraphics[height=0.8cm]{int2.pdf}}
  \end{picture} \hskip 1.7cm \ + \sum_{m,n=0}^\infty \ \ \ \ \ \ \ \ \
  \ \ \ \ \setlength{\unitlength}{3pt} \begin{picture}(-15, -15)(9,5)
  \put(-3.5, +2){\includegraphics[height=0.8cm]{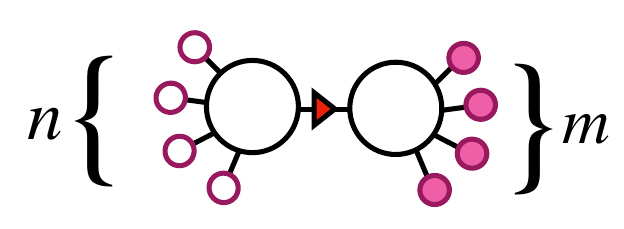}}
  \end{picture} \hskip 1.7cm \ \ \ \ \ \ \ \ \ \ \ \ +
  \sum_{m,n=0}^\infty \ \ \ \ \ \ \ \ \ \ \ \ \
  \setlength{\unitlength}{3pt} \begin{picture}(-15, -15)(9,5)
  \put(-3.5, +2){\includegraphics[height=0.8cm]{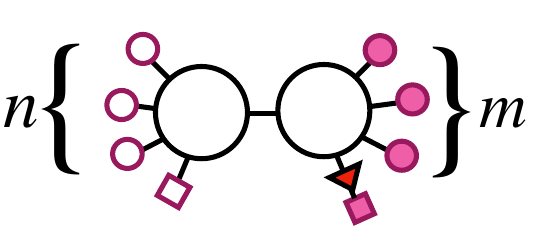}}
  \end{picture} \hskip 1.7cm
  \eee
The diagram expansion comprises diagrams with the structure of
trees with arbitrary number of cycles.  The sum over trees can be
absorbed into the definition of the vertices if the Feynman rules are
determined by the expansion around the mean field solution
\re{Widentityator} .

   \footnotesize

\providecommand{\href}[2]{#2}\begingroup\raggedright\endgroup

\end{document}